\documentclass[11pt, letterpaper]{article}

\usepackage[margin=1in]{geometry}
\usepackage{setspace}
\usepackage[noblocks]{authblk} 

\usepackage{tgtermes}
\usepackage{newtxtext}
\usepackage{newtxmath}
\usepackage{bm}
\usepackage{array}

\usepackage{algorithm}
\usepackage{algpseudocode}
\usepackage{tikz}
\usepackage{booktabs}
\usepackage{graphicx}

\usepackage{amsthm}

\usepackage[numbers]{natbib}
 \bibpunct[, ]{(}{)}{,}{a}{}{,}%
\usepackage[colorlinks=true, linkcolor=blue, citecolor=blue, urlcolor=blue]{hyperref}

\title{The hybrid confirmation tree: A robust strategy for hybrid intelligence}

\author[1]{Julian Berger}
\author[2]{Pantelis P. Analytis}
\author[3]{Frederik Andersen}
\author[4]{Kristian P. Lorenzen}
\author[5]{Ville Satopää}
\author[1]{Ralf HJM Kurvers}

\affil[1]{Max Planck Institute for Human Development}
\affil[2]{University of Southern Denmark}
\affil[3]{PwC}
\affil[4]{Danske Commodities}
\affil[5]{INSEAD}

\date{\today}

\begin{document}

\maketitle

\begin{abstract}
Combining human and artificial intelligence (AI) is a potentially powerful approach to boost decision accuracy. However, few such approaches exist that effectively integrate both types of intelligence while maintaining human agency. Here, we introduce and evaluate the hybrid confirmation tree, a simple aggregation strategy that compares the independent decisions of both a human and AI, with disagreements triggering a second human tiebreaker. Through analytical derivations, we show that the hybrid confirmation tree can match and exceed the accuracy of a three-person human majority vote while requiring fewer human inputs, particularly when AI accuracy is comparable to or exceeds human accuracy. We analytically demonstrate that the hybrid confirmation tree's ability to achieve complementarity---outperforming individual humans, AI, and the majority vote---is maximized when human and AI accuracies are similar and their decisions are not overly correlated. Empirical reanalysis of six real-world datasets (covering skin cancer diagnosis, deepfake detection, geopolitical forecasting, and criminal rearrest) validates these findings, showing that the hybrid confirmation tree improves accuracy over the majority vote by up to 10 percentage points while reducing the cost of decision making by 28--44$\%$. Furthermore, the hybrid confirmation tree provides greater flexibility in navigating true and false positive trade-offs compared to fixed human-only heuristics like hierarchies and polyarchies. The hybrid confirmation tree emerges as a practical, efficient, and robust strategy for hybrid collective intelligence that maintains human agency.
\end{abstract}

\section*{Introduction}

Research on harnessing the wisdom of crowds has produced a diverse set of strategies for aggregating individual judgments into a more accurate group decision \citep{surowiecki2005wisdom, kurvers2016boosting, kurvers2023automating, budescu2015identifying, mannes2014wisdom}. While human-only crowds can offer substantial benefits over individual decision makers \citep{kurvers2016boosting, surowiecki2005wisdom, mannes2014wisdom}, the recent leap in the performance of artificial intelligence (AI) presents a compelling opportunity: Is it possible to design effective and cost-efficient hybrid systems that combine human and algorithmic competences? This question is particularly relevant in domains where human expertise is valuable but scarce, and where AI may offer complementary strength but cannot be trusted as a standalone solution.

Traditional approaches to group decision making often involve robust strategies like majority voting \citep{hastie2005robust, csaszar2013organizational} or more structured organizational forms such as hierarchies or polyarchies, where a single individual's decision can determine the collective outcome \citep{sah1986architecture,christensen2010design}. Although these strategies can substantially improve performance compared to individual decision making, they often require several individuals, and their performance can be hindered by correlated errors within human groups \citep{kurvers2019detect, palley2019extracting, herzog2019ecological, clemen1985limits}. As AI models become increasingly competent across various decision-making problems \citep{brinker2019comparing, csaszar2024artificial, karger2025forecastbench}, sometimes performing at or even beyond the level of human experts, the need to study decision-making structures that combine humans and artificial agents is gaining urgency, given that fully autonomous AI systems may lack the nuanced understanding, ethical grounding, or public trust afforded to human decision makers, particularly in high-stakes scenarios \citep{rahwan2019machine, santoni2021four, bonnefon2024moral}. 

\begin{figure}[!t]
    \centering
    \includegraphics[width = \textwidth]{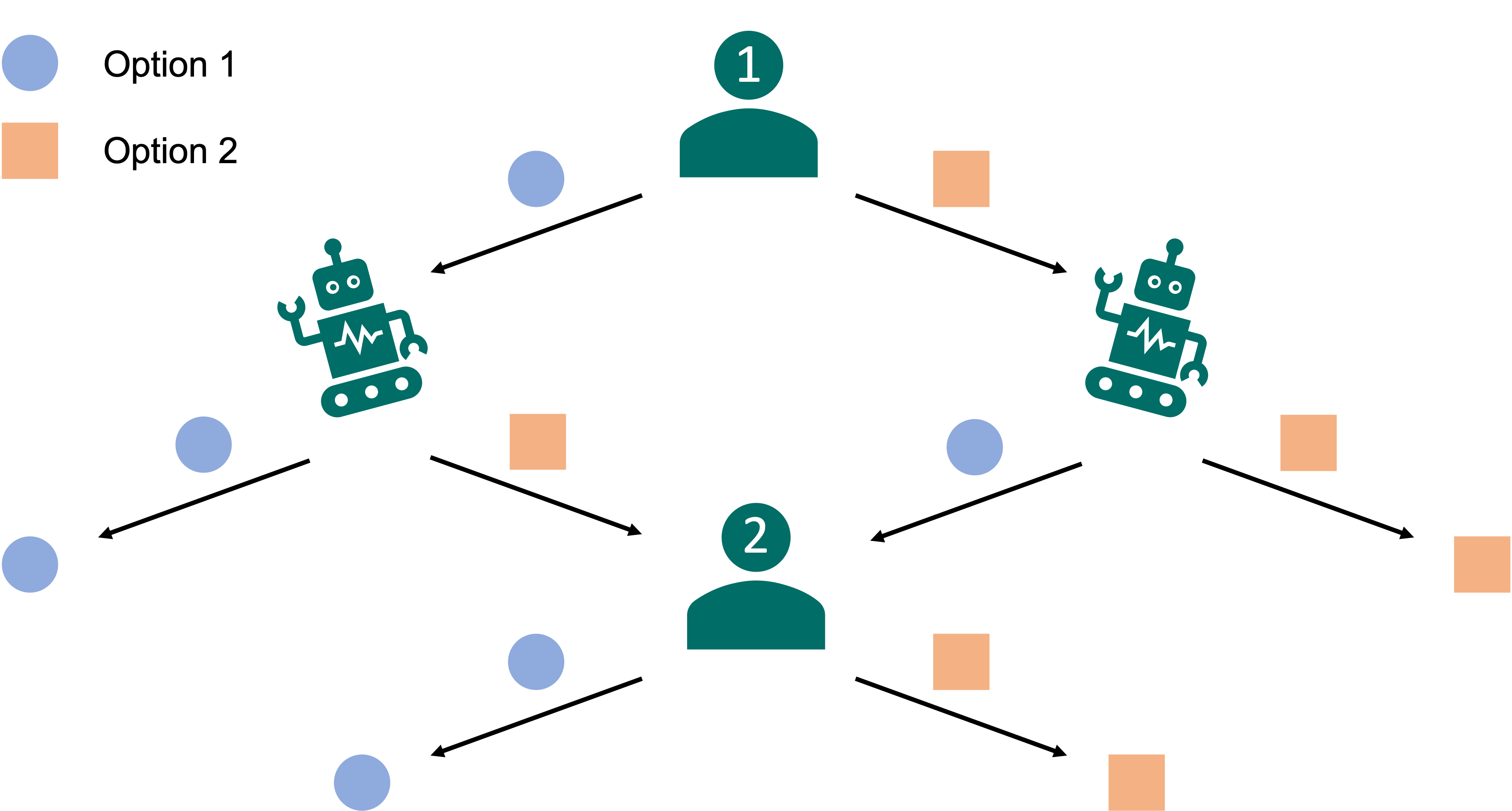}
    \caption{The hybrid confirmation tree procedure. The independent decisions of a human and AI are compared. In cases of agreement, that decision is accepted. In cases of disagreement, a second human breaks the tie. The final decision is therefore always approved by at least one human.}
    \label{tree}
\end{figure}

This paper evaluates the hybrid confirmation tree (Figure \ref{tree}), a simple and transparent strategy that combines human expertise and artificial intelligence. The hybrid confirmation tree operates sequentially: An initial human decision is compared to that of a machine and if they agree, that decision is accepted. If not, a second human breaks the tie, providing a final resolution. In either case, a human always approves the final decision. This structure offers a pragmatic approach to integrating machines into human workflows, aiming to harness the wisdom of hybrid crowds by building on the potential for AI to be more accurate or efficient than a human, while retaining human oversight and reducing the overall human effort compared to human-only aggregation strategies.

Combining analytical derivations varying human and AI accuracies and their correlations with empirical analysis of six real-world datasets from the domains of medical diagnostics, deepfake classification, geopolitical forecasting, and criminal rearrest predictions, we demonstrate the conditions under which the hybrid confirmation tree excels. We show that the hybrid confirmation tree is at least as good as the human majority vote but more efficient as long as the machine's decisions are as good as humans' decisions, even when human and machine decisions are correlated. Empirically, we find that the hybrid confirmation tree is more accurate than the human majority vote by up to 10 percentage points while saving anywhere between 28$\%$ and 44$\%$ of human choices. Furthermore, we show that the hybrid confirmation tree can navigate the inherent errors of true and false positives more flexibly than simpler two-person structures such as hierarchies (which produce few positive classifications, as both humans need to agree on a positive classification) and polyarchies (which produce many positive classifications, as one human can settle the final decision with a positive classification). Our findings highlight the potential of the hybrid confirmation tree as a practical heuristic for achieving superior decision accuracy at a reduced cost compared to the human-only majority vote, while also offering insights into the critical role of human--AI correlation in shaping the performance of hybrid intelligence systems.

\section*{Related Literature}
\subsection*{Wisdom of crowds}

The wisdom of crowds literature demonstrates that aggregating independent human judgments can substantially boost decision accuracy compared to individual decision making \citep{surowiecki2005wisdom, kurvers2016boosting, kurvers2023automating}. A quintessential aggregation rule in binary choice is the majority vote, which selects the option favored by the majority. The effectiveness of majority voting depends on two conditions: Individual accuracy must exceed chance levels, and decision makers' errors should not be highly correlated \citep{de2014essai, grofman1983thirteen, herzog2019ecological}. When these conditions are met, crowds generally achieve higher accuracy, since individual errors cancel out in aggregation. 

However, the assumption of statistically independent choices between decision makers is often violated in real-world settings. Even when decision makers make independent decisions, they tend to rely on similar information or process available information similarly \citep{bikhchandani2000herd, gigerenzer_heuristic_2011, veale2017visual, palley2019extracting}, which may lead to correlated decisions. Even worse, social influence can further amplify correlated errors that undermine the wisdom of crowds \citep{lorenz2011social, clemen1985limits, grofman1983thirteen}. This problem is particularly acute in sequential decision making, where early opinions can disproportionately influence later ones, creating informational cascades that lock the group into incorrect choices \citep{frey2021social, lorenz2011social}. The challenge of correlated decisions thus highlights the importance of designing aggregation processes that either preserve the independence of judgments or are robust to its absence.

While a variety of aggregation strategies exist beyond simple majority voting---such as weighting schemes that use past performance to select or upweight successful individuals \citep{budescu2015identifying, mannes2014wisdom}---these methods often require historical performance data. In absence of such data, decision similarity can be a successful proxy of decision accuracy \citep{kurvers2016boosting, himmelstein2023wisdom}, since more accurate decision makers will more frequently agree on correct choices. Agreement between decision makers can therefore be seen as indicating a correct answer as long as individual-level accuracy is above chance. Moreover, simple, unweighted majority voting often performs remarkably well across a wide range of statistical environments and is surprisingly difficult to consistently outperform, even with more sophisticated aggregation methods that rely on weighting \citep{hastie2005robust}. This robustness stems from its simplicity: The majority rule requires no parameter estimation and is therefore not prone to overfitting on noisy data (see also \citet{herzog2019ecological} and \citet{analytis2018social} for discussions of the bias--variance dilemma in crowd wisdom). 

The hybrid confirmation tree integrates three central facilitators of successful crowd aggregation: It relies on independent choices, it finalizes a choice whenever there is agreement between (human and machine) decision makers, and it relies on simple voting in the form of the majority vote. 

\subsection*{Organizational decision making}

\citet{sah1986architecture, sah1988committees} established the theoretical framework for understanding how organizational structures manage different types of decision errors, demonstrating that organizational decision-making structures must navigate trade-offs between false positives (accepting bad proposals) and false negatives (rejecting good proposals). Their analysis of hierarchies and polyarchies revealed principles that continue to shape organizational design theory. In hierarchies, proposals must receive unanimous approval from sequential decision makers, creating a conservative system that filters out bad decisions but may also reject good ones. Polyarchies operate through parallel decisions where only a single approval is needed, creating a more liberal system that accepts more good proposals but potentially also more bad proposals. The optimal choice between these structures depends on the relative costs of false positives and false negatives in the specific decision environment \citep{sah1988committees}. 

The hybrid confirmation tree offers a new approach to navigating these fundamental trade-offs by leveraging a key advantage of machine learning: the ability to specify how machines should make errors. While human decision makers may be difficult to train for specific error profiles and may be inconsistent in their application of decision criteria, machine learning models can be specified for the trade-off between false positives and false negatives. Including a model that favors one error over the other in the hybrid confirmation tree can therefore help to navigate the trade-off. This approach aligns with the work of \citet{christensen2010design} on intermediate organizational forms that achieve error trade-offs unattainable by pure hierarchies or polyarchies, but at the cost of consulting more decision makers. The hybrid confirmation tree extends this insight by showing how human--AI combinations can achieve similar benefits at lower costs.

Organizational decision-making structures fundamentally shape organizations' search and learning processes. \citet{knudsen2007two} demonstrated that the choice between hierarchical and polyarchical structures directly influences an organization's capacity for exploration and exploitation: Hierarchies tend to promote exploitation over exploration, since new options are rarely acted on due to the need for unanimous approval, whereas the opposite is the case for polyarchies, since exploring new options only requires the approval of one decision maker. \citet{csaszar2013organizational} formalized this relationship by showing how different organizational structures create distinct frontiers in the exploration--exploitation space: They can optimize for either exploration or exploitation by selecting the threshold for a new option to be explored, again at the cost of increasing the number of decision makers and the size of the quorum needed to engage in a new option. The hybrid confirmation tree contributes to this discussion: While its decision-making structure remains the same (Figure \ref{tree}), the flexibility of the machine in the decision-making structure makes it possible to explore more or fewer options depending on the machine's setup.

\subsection*{Human and AI ensembles}

The integration of AI into organizational decision making presents new opportunities and challenges for information aggregation. \citet{csaszar2024artificial} argued that AI can extend the bounds of human decision making by serving as an additional information source that complements human judgment. However, a key consideration is how to structure human--AI interaction so as to maximize these benefits while avoiding the pitfalls of relying on the machine even when it is wrong. The dominant paradigm treats AI as an advisor that provides information to influence human deliberation \citep{glikson2020human}. This approach faces challenges in calibrating human reliance on machines appropriately. Research in human--computer interaction has shown that people trust bad AI choices \citep{bansal_does_2021, bucinca_trust_2021} and fail to trust good AI choices \citep{bansal_does_2021, chiang2023two}. To alleviate this pattern of over- and underreliance, researchers had humans make choices independent of machines first, then allowed them to revise their choice after seeing AI advice \citep{yin2025designing, buccinca2021trust}. Meta-analytic evidence reveals that human--AI combinations often do not perform as well as the better-performing partner of the pair. In other words, a human--AI combination that outperforms both human and machine accuracy is difficult to achieve \citep{vaccaro2024combinations}. \citet{puranam2021human} proposed an alternative approach in which humans and AI tackle the same decision task independently and their judgments are then aggregated. This framework treats human--AI collaboration as an aggregation problem similar to traditional crowd wisdom, but with the key advantage that humans and AI may access different types of information. \citet{choudhary2025human} formalized this aggregation problem, distinguishing between digital data accessible to AI and experiential knowledge available only to humans. They argued that human--AI combinations are most likely to achieve complementary performance when both types of information are valuable for the decision problem---thereby  hypothesizing that aggregation success rests on uncorrelated decisions between humans and machines. Empirically, we find that the success of independent human--AI aggregation appears to depend on uncorrelated decisions. The extent to which humans and AI make different types of mistakes helps to compensate for one party's errors. When human and AI errors are uncorrelated, aggregation can achieve the error-canceling benefits observed in traditional human-only crowd wisdom \citep{steyvers_bayesian_2022, zoller2025human}.

Despite this theoretical promise, practical implementations of independent human--AI aggregation remain limited. Most existing approaches require complex weighting schemes and substantial training data, as is the case for the Bayesian graphical model by \citet{steyvers_bayesian_2022} and the weighted ensemble by \citet{zoller2025human}. This gap suggests an opportunity for simpler aggregation strategies---such as the hybrid confirmation tree's use of a majority vote between AI and humans---that can achieve the benefits of hybrid intelligence while remaining feasible for real-world organizational settings with limited computational resources and data availability.

\section*{Results}

\subsection*{The intuitive benefit of hybrid confirmation trees }

Hybrid confirmation trees are a simple heuristic for arbitrating between human and machine choices and can be represented in a decision tree structure (Figure \ref{tree}). First, a human and a machine decide independently. If they agree, that option is accepted. If they disagree, a second human comes in to break the tie.

We start by comparing the hybrid confirmation tree to its closest relative, the human majority vote. The hybrid confirmation tree resembles a three-person sequential majority vote in which the second human is replaced by a machine. An analytical comparison between both approaches (see Methods for analytical derivatives) showed two distinct effects. First, when the machine performance is higher than the average human performance, the hybrid confirmation tree will outperform the majority vote, and when the machine performance is lower than the average human performance, the hybrid confirmation tree will perform worse than the majority vote. Second, the hybrid confirmation tree is more cost-effective, as it requires substantially fewer ratings than the majority vote. Even if machine and human performance are identical and hence both the hybrid confirmation tree and the majority vote are equally accurate, the hybrid confirmation tree can substantially reduce costs. 

\subsection*{Correlated decisions hinder complementary performance}

\begin{figure}[!t]
    \centering
    \includegraphics[width = \textwidth]{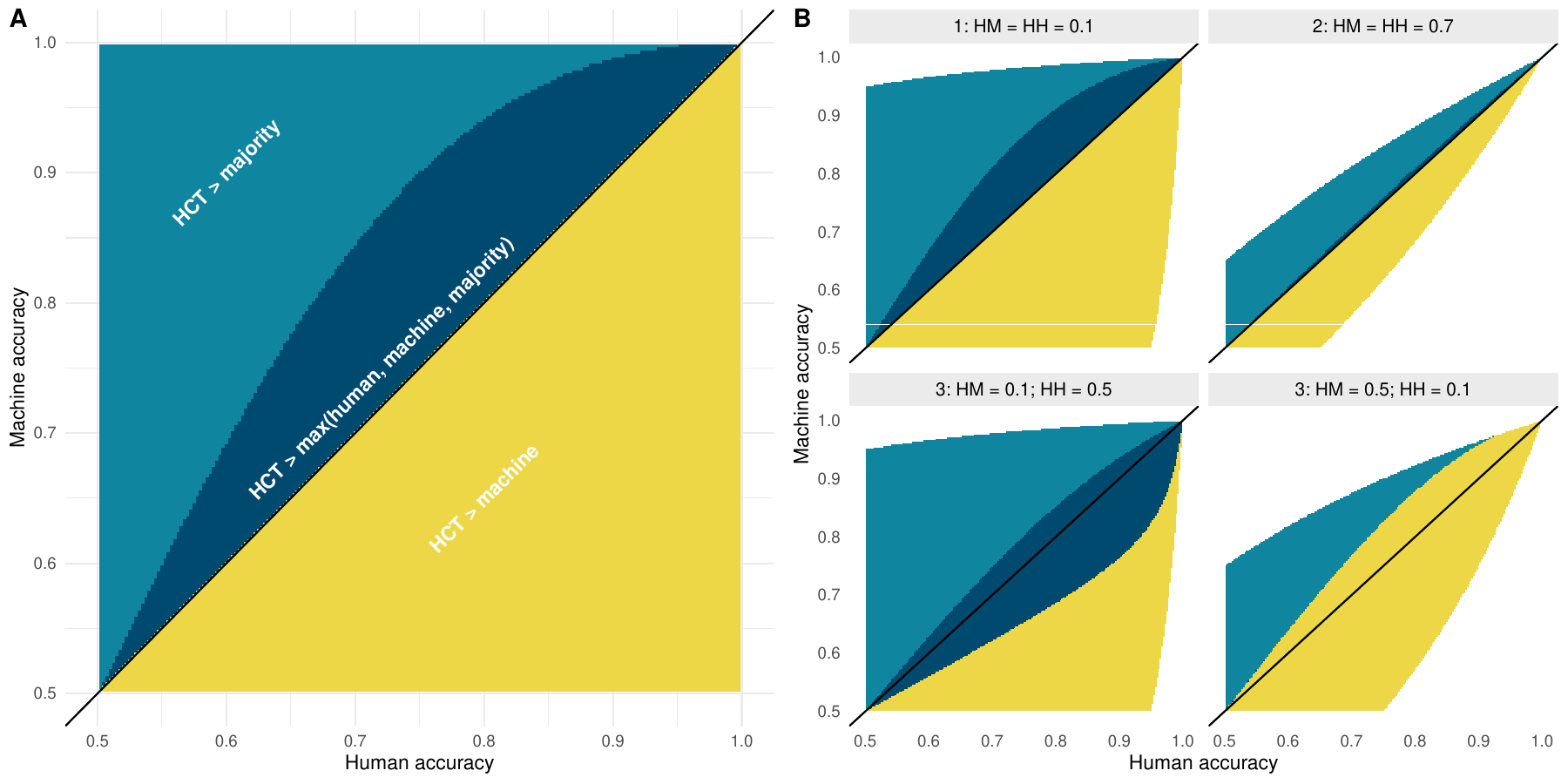}
    \caption{Comparing the accuracy of the hybrid confirmation tree, a machine alone or humans alone, and a three-person majority vote as a function of human accuracy and machine accuracy. (A) Regions where the hybrid confirmation tree outperforms only the majority vote (blue), only the machine (yellow), and both the majority vote and the machine (dark blue). (B) Results shown in (A) but for varying levels of human--machine (HM; $\kappa_{HM}$) and human--human (HH; $\kappa_{HH}$) dependence. Note that high $\kappa$-values are only possible for more similar human and machine accuracies.}
    \label{fig3}
\end{figure}

These results suggest that the hybrid confirmation tree can be a powerful substitute for human-only decision making. Next, we mapped out the conditions under which the hybrid confirmation tree can outperform the majority vote, AI, or both, by simulating different levels of human and machine accuracies (see Methods for details). Figure \ref{fig3}A shows that when human accuracy was higher than machine accuracy, the hybrid confirmation tree outperformed the machine (but not the majority vote), and when machine accuracy was higher than human accuracy, the hybrid confirmation tree outperformed the majority vote. There was one region of interest in which the hybrid confirmation tree was able to outperform both the machine and the majority vote: when both humans and machines were better than chance, and machine accuracy was slightly higher than human accuracy. Supplementary Figure \ref{si_full_theory} shows that in the unlikely event that both human and machine accuracy are worse than chance and humans are slightly better than machines, the hybrid confirmation tree performs worse than the majority vote and the machine.

Thus far, we have assumed that choices between humans and between humans and machines are statistically independent. Yet correlated decisions are known to inhibit collective success \citep{grofman1983thirteen, clemen1985limits, herzog2019ecological, palley2019extracting}. Moreover, recent research shows that human decisions tend to be more correlated with other human decisions than with machine decisions \citep{zoller2025human, steyvers_bayesian_2022}. To investigate the effect of correlation in decisions among humans, and between humans and machines, we modeled the success of the hybrid confirmation tree and the human majority vote with various levels of decision correlation. To vary the level of correlation, we used Cohen's $\kappa$, combined with an opinion-leader approach \citep{grofman1983thirteen, kurvers2019detect} (see Methods for analytical derivatives). For the hybrid confirmation tree, we varied how much the machine agreed with the human leader ($\kappa_{HM}$) and how much the second human agreed with the human leader ($\kappa_{HH}$). For the human majority vote, we applied this same opinion-leader structure by designating one human as the leader and the other two as followers, setting the correlation between the leader and each follower to same human--human correlation ($\kappa_{HH}$).

Figure \ref{fig3}B illustrates four prototypical scenarios of human--human and human--machine correlation, as well as their effect on the potential for complementary performance. When human--human and human--machine correlations are low (top-left panel), a similar pattern emerges as when assuming complete independence (Figure \ref{fig3}A). When both correlations are high (top-right panel)---that is, when humans and machines make near-identical decisions---there is no region in which the hybrid confirmation tree can outperform both the majority vote and the machine. In other words, increasing the correlation among decisions reduces the scope for complementary performance. When both correlations are high, the deciding factor is the relative accuracy of the machine compared to the humans: When the machine is better than humans, the hybrid confirmation tree will be better than the majority vote but still less accurate than the machine alone.

The bottom panels show different levels of correlation between humans and between humans and machines. When human--human correlation is high and human--machine correlation is low (bottom-left panel), the region in which the hybrid confirmation tree outperforms both the majority vote and the machine encroaches into the territory where the human majority vote performs best under the assumption of independence (i.e., below the diagonal). The high dependence between human choices allows the hybrid confirmation tree to outperform the human majority vote in this region, despite the machine performing worse than human accuracy. At the same time, the area for complementary performance when the machine is better than humans (above the diagonal) is reduced compared to independent choices (Figure \ref{fig3}A) and lower human--human dependence (Figure \ref{fig3}B, top-left panel). In the highly unlikely scenario that human--machine correlation is high and human--human correlation is low (bottom-right panel), there is no region in which the hybrid confirmation tree outperforms both the majority vote and the machine, and the majority vote encroaches into the region of the hybrid confirmation tree. Supplementary Figure \ref{si_kappa} shows results for a broader range of $\kappa_{HM}$ and $\kappa_{HH}$ values.

In sum, our analytical results suggest that the hybrid confirmation tree is a promising alternative to the human majority vote as long as machine accuracy is similar to or better than average human individual accuracy, and human--machine decisions are less correlated than human--human decisions.

\subsection*{Improving decision making in high-stakes domains}

We next tested the performance of the hybrid confirmation tree against the human majority vote in six previously published datasets containing both human and machine choices: one dataset on the prediction of criminal rearrest, one dataset on the detection of deepfake videos, two datasets on geopolitical forecasting and 
two datasets on skin cancer diagnoses, one based on dermoscopic images and one on nondermoscopic images. For three datasets (detecting deepfakes and making geopolitical forecasts) we relied on an existing machine algorithm; for the other three we trained our own classifier (see Table \ref{tab:summary} for a summary of the datasets and Methods for a description of each dataset).

\newcolumntype{L}[1]{>{\raggedright\arraybackslash}p{#1}}
\begin{table}[t]
\footnotesize
\centering
\caption{Summary of datasets: Number of cases, human raters, total human choices, and AI model type per domain.}
\begin{tabular}{@{} 
    >{\bfseries}L{0.3\textwidth}   
    >{}L{0.2\textwidth}     
    r                               
    r                               
    r                               
    L{0.25\textwidth}               
  @{}}
\toprule
Domain & Citation & Cases & Humans & Choices & Type of machine (Source) \\
\midrule
Skin Cancer (Derm)        & \citet{brinker2019deep,brinker2019comparing}   &   100  &   157  &  15,700 & CNN (own model)                                 \\
Skin Cancer (Nonderm)     & \citet{brinker2019deep,brinker2019comparing}   &   100  &   145  &  14,500 & CNN (own model)                                 \\
Deepfakes          & \citet{groh2022deepfake}                       &    54  &   132  &   1,347 & CNN \citep{seferbekov2021deepfake}               \\
Criminal Rearrest & \citet{angwin2016machine,dressel2018accuracy}  & 1,000 &   400  &  20,000 & Logistic regression (own model)                  \\
Hybrid Forecasting Competition & \citet{benjamin2023hybrid}  &    52  &   111  &   1,055 & Time series regression \citep{benjamin2023hybrid} \\
ForecastBench      & \citet{karger2025forecastbench}                &   422  &   500  &  21,302 & LLM \citep{karger2025forecastbench}              \\
\bottomrule
\end{tabular}
{\raggedright CNN stands for convolutional neural network, LLM stands for large language model. \par}
\label{tab:summary}
\end{table}

We implemented the hybrid confirmation tree in the following way: For each case in each dataset, we generated all pairwise permutations between two humans and inserted the machine choice in the middle. Though human choices were provided as binary choices, machine choices are frequently provided as probability predictions between 0 or 1 (i.e., the probability with which the machine deems a case to belong to the positive class). These probabilities are turned into a binary choice by selecting a threshold that binarizes anything above it as a positive class and anything below it as a negative class. We tested the performance of the hybrid confirmation tree for all thresholds between 0 and 1. We determined these thresholds by examining the unique values a model returned on a given dataset; for instance, if a model only returned three distinct predictions, we tested only those three for binarizing choices into positives and negatives. Next, for every pairwise permutation, we recorded whether the first human and the machine agreed in their choice. If they did, that choice was taken as the final choice of the hybrid confirmation tree and the cost was set to 1, as only one human was consulted. If the first human and the machine disagreed, we took the choice of the second human as the final choice of the hybrid confirmation tree and set the cost to 2. We then determined whether the hybrid confirmation tree was correct or not by comparing the final choice to the correct answer (all six datasets contained a ground truth; see Methods for details). Finally, we averaged the frequency of correct predictions and the cost within a case over all permutations of human and machine triplets and averaged across cases to achieve a general measure of accuracy and cost per dataset and threshold. We proceeded similarly with the human majority vote, generating up to 25,000 unique permutations of three decision makers for each case (if fewer than 31 humans were available for a case, we used the maximum number of permutations possible) and determining the majority outcome, its accuracy, and the frequency of agreement between the first two decision makers (to determine the costs of the human majority vote). Similarly, we first averaged results within cases, then across cases within datasets, to determine the accuracy and costs per dataset. 

\begin{figure}[!t]
    \centering
    \includegraphics[width = \textwidth]{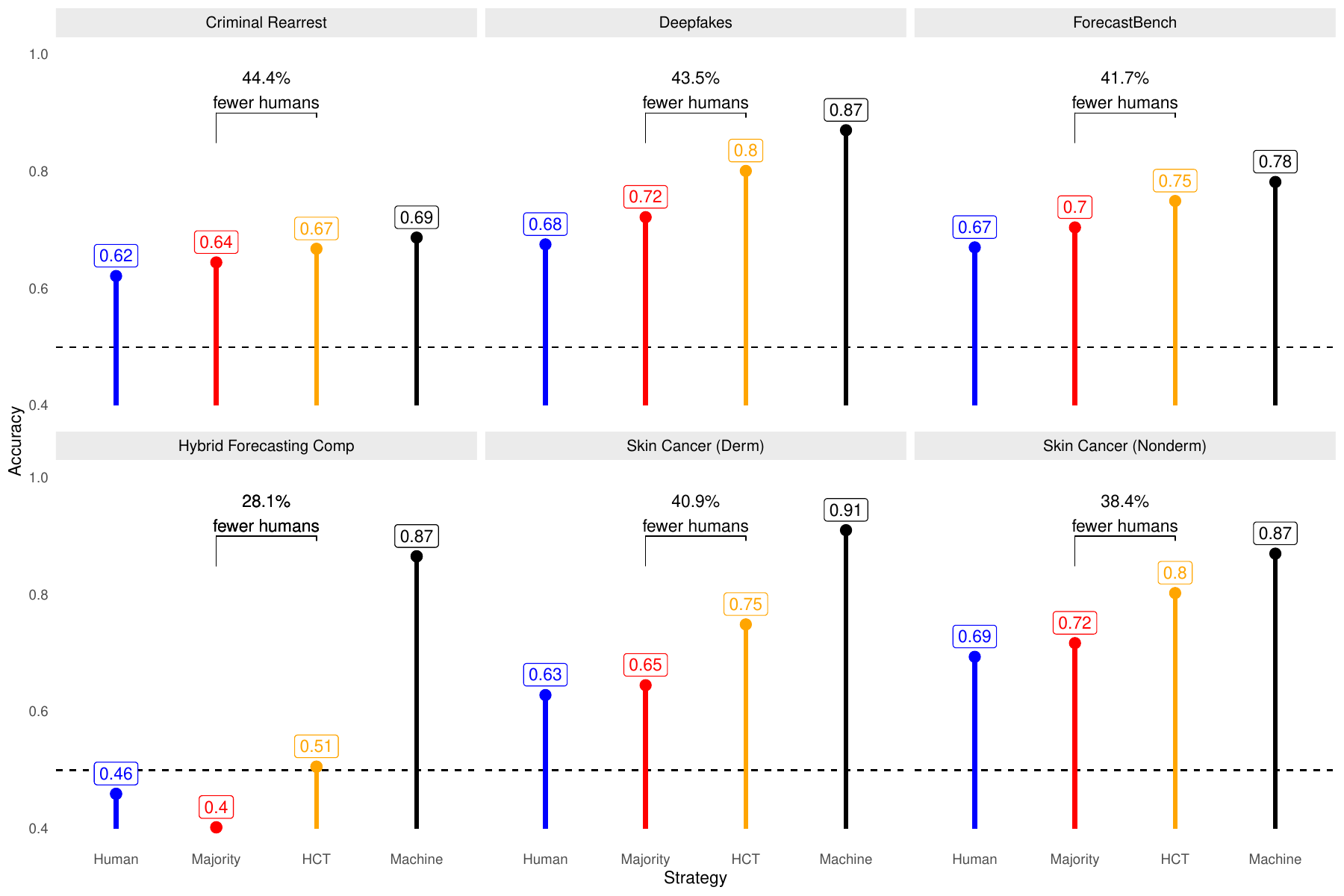}
    \caption{The accuracy of single humans, a three-person majority vote, the hybrid confirmation tree (HCT), and a machine, per domain. Results were generated by selecting the threshold setting that maximized hybrid confirmation tree accuracy. Dashed horizontal lines indicate chance level (i.e., accuracy of 0.5). Brackets indicate the relative reduction in human decision makers of the hybrid confirmation tree compared to the human majority vote.}
    \label{fig4}
\end{figure}

Figure \ref{fig4} presents the results for the threshold setting that maximizes the accuracy of the hybrid confirmation tree. The hybrid confirmation tree improved on the human majority vote in all domains. We quantified the statistical credibility of the accuracy difference between the hybrid confirmation tree and the majority vote per domain using Bayesian estimation (see Methods for details). The hybrid confirmation tree credibly improved accuracy in all datasets by at least 2.3 percentage points (highest density interval [HDI] 1.4 to 3.2) in the criminal rearrest data and by as much as 10.4 percentage points (HDI 7.9 to 12.9) in the skin cancer diagnoses based on dermoscopic images. Supplementary Table \ref{tab:posterior} presents a detailed accuracy comparison for every domain. The hybrid confirmation tree reduced the required number of human decision makers by between 28$\%$ (Hybrid Forecasting Competition) and 44$\%$ (criminal rearrest) compared to the majority vote. Next, we tested whether the selection of the accuracy-maximizing threshold was possible out of sample by running a repeated cross-validation procedure (see Methods for details), selecting the threshold that maximized accuracy in a training set and applying it to new cases in a test set. Supplementary Figure \ref{si_cv_marginals} shows that the accuracy differences between all strategies observed in Figure \ref{fig4} remained stable when selecting the threshold out of sample. Supplementary Figure \ref{si_larger_crowds} compares the performance of the hybrid confirmation tree against the majority vote for larger crowds, showing that the hybrid confirmation tree outperformed even large crowds of 15 individuals in all six datasets.

Although the hybrid confirmation tree consistently outperformed the majority vote (even for larger crowds), it consistently underperformed compared to the machine alone (even when the machine was only slightly more accurate than humans alone, as was the case in the criminal rearrest dataset). To better understand this result, and to determine the match between our empirical and theoretical results, we calculated the human--human and human--machine correlations (i.e., Cohen’s $\kappa$) for every possible combination of decision makers in each dataset (see Methods for details). Unsurprisingly, both human--human and human--machine agreement were above the level expected by their accuracies alone (i.e., $\kappa > 0$; Supplementary Figure \ref{si_kappa_empirical}). In five domains, human--human correlation was higher than human--machine correlation (median difference ranged from 0.09 in the deepfakes dataset to 0.704 in the Hybrid Forecasting Competition dataset). In the criminal rearrest dataset, human--machine correlation was slightly higher than human--human correlation (median difference: 0.034). Supplementary Figure \ref{si_theory_data} redraws Figure \ref{fig3} for each dataset based on their observed human--human and human--machine correlations, showing that all datasets fell in the area in which the hybrid confirmation tree is predicted to outperform the human majority vote but not the machine, suggesting a good match between the theoretical and empirical results.

\subsection*{Hybrid confirmation trees make flexible error trade-offs}

\begin{figure}[!h]
    \centering
    \includegraphics[width = \textwidth]{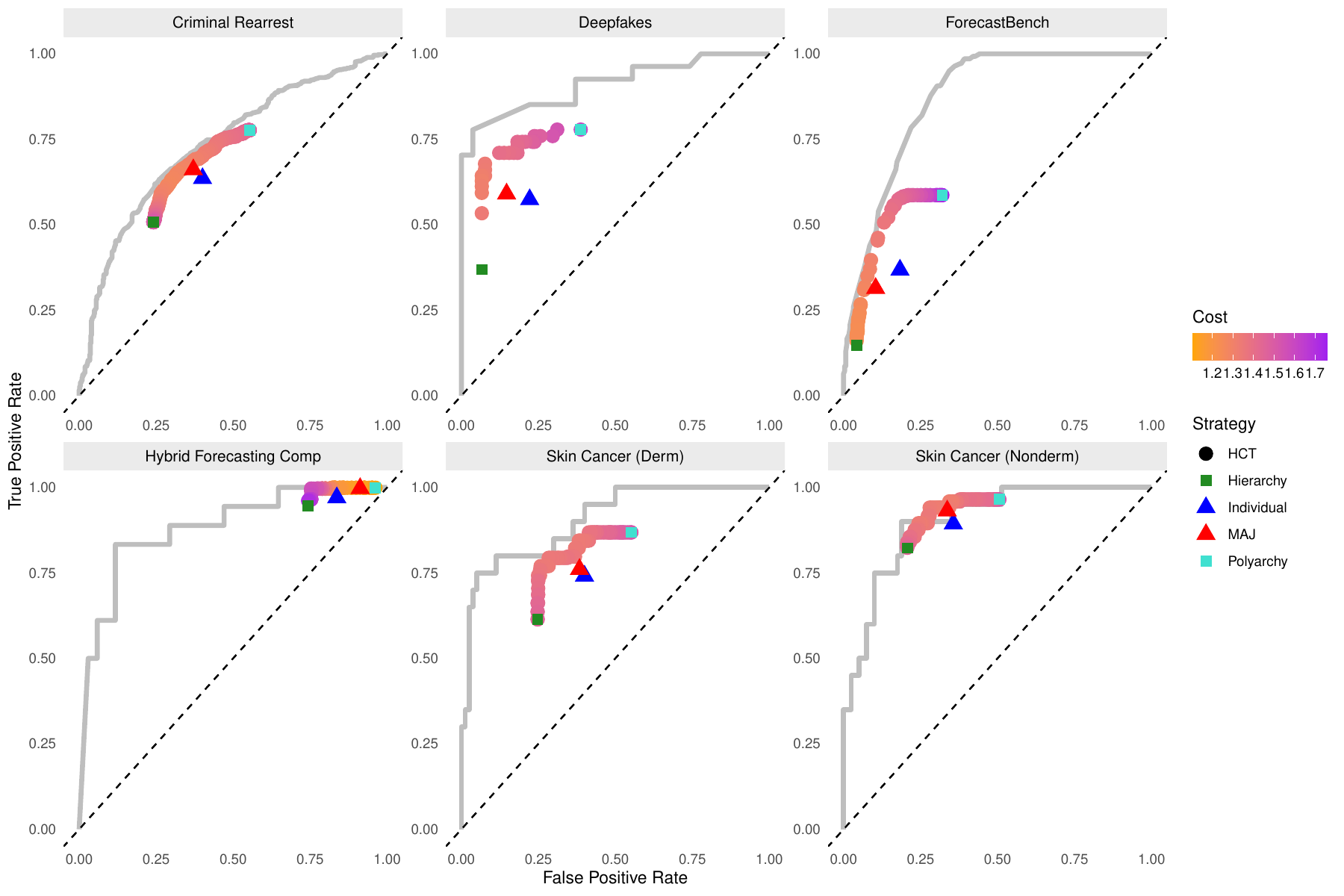}
    \caption{The hybrid confirmation tree achieved flexible error trade-offs across domains. Colored dots show the true positive (y-axis) and false positive rates (x-axis) for the hybrid confirmation tree (HCT) for thresholds between 0 and 1, governing the likelihood that the machine learning model counts a case as positive. Color coding corresponds to the cost of the hybrid confirmation tree in terms of the average number of human decision makers needed to make a decision. The performance of the average human decision maker (blue triangle), the human majority vote (MAJ; red triangle), the hierarchy (green square) and the polyarchy (turquoise square) are also shown. The gray line denotes the machine's receiver operator characteristic (ROC) curve.}
    \label{fig5}
\end{figure}

Because machines provide probability predictions between 0 and 1, one advantage of the hybrid confirmation tree over the human majority vote is that it allows for direct arbitration between true and false positives by setting different threshold values. When the threshold is set to 0, the machine will label every image as a positive case, meaning that both humans would need to label it as a negative case in order for the hybrid confirmation tree to override the machine. When the threshold is set to 1, both humans would need to provide a positive choice for the case to be labeled positive. This threshold setting thus directly governs the trade-off between true and false positives and can be used to align with a preference to be conservative and avoid false positives or a preference to be lenient and avoid false negatives. For example, avoiding costly false positives may be a priority for investments, whereas in cancer screenings, a high rate of false positives may be acceptable if the cost of missing a malignant tumor is high.

There are close parallels with human-only organizational structures for arbitrating between choices, namely hierarchies and polyarchies \citep{sah1986architecture, sah1988committees, christensen2010design}. Consider two medical professionals aiming to identify whether a tumor is benign or malignant. In hierarchies of two, both professionals need to agree that an image is malignant in order to classify it as such. In polyarchies of two, only one professional needs to classify the image as malignant; agreement is not necessary. A threshold value of 1 in the hybrid confirmation tree makes it behave like a hierarchy of two, as both humans in the hybrid confirmation tree need to decide on the positive class for that to be the final outcome. Likewise, a threshold value of 0 makes the hybrid behave like a two-person polyarchy, since a single positive classification made by a human is sufficient to classify a case as positive. Figure \ref{fig5} shows the true positive and false positive rates for each threshold for the hybrid confirmation tree (as well as the machine performance for different thresholds) in the receiver operating characteristic (ROC) space. The hybrid confirmation tree is able to operate on a range of true positive and false positive trade-offs depending on the threshold used for the machine classification. This is in contrast to the human majority vote, which has just one result in the ROC space. The hierarchy and polyarchy also each have just a single result in the ROC space, both equal to what the hybrid confirmation tree achieves at its extremes (i.e., thresholds of 0 and 1). Across the domains, the hybrid confirmation tree reproduces the shape of the machine’s ROC curve (gray line) but at a reduced performance, as the hybrid confirmation tree underperforms compared to the machine alone.

The hybrid confirmation tree thus allows for a more flexible error trade-off compared to human-only strategies. Note that in every domain there is at least one point of the hybrid confirmation tree performance (i.e., one threshold setting of the classification algorithm) that improves both true positive and false positive rates compared to the human majority vote (see Supplementary Figure \ref{si_zoom} for a more detailed view). For instance, the top-left points of the hybrid confirmation tree in the deepfake dataset are better than the human majority vote (red triangle) in both correctly identifying deepfakes and avoiding incorrectly labeling videos without deepfakes. 

The optimal threshold depends on subject matter experts' goals and preferences. Supplementary Figure \ref{si_cost} shows the performance of the hybrid confirmation tree when the goal is to minimize human cost by selecting the threshold that minimizes how often the hybrid confirmation tree consults a second human. Supplementary Figure \ref{si_auc} shows the results when maximizing the area under the ROC curve (AUC). The hybrid confirmation tree thus allows decision makers to navigate the trade-off between true and false positives much more flexibly than the human majority vote.

\section*{Discussion}

The hybrid confirmation tree combines human and machine choices and is consistently able to improve upon human-only baselines---namely, a human majority vote between three people and two-person hierarchies and polyarchies---across analytical derivations and empirical investigations of six high-stakes datasets, outperforming even much larger human crowds. The hybrid confirmation tree is also more flexible in trading off true positives and false positives, as the decision threshold of the implemented algorithm can be adjusted according to the costs associated with different types of categorization errors. Importantly, the hybrid confirmation tree is a reliable way to reduce workload in domains where human experts are scarce: It achieves better or equivalent true positive and false positive rates compared to the human majority vote as well as to hierarchies and polyarchies, while saving anywhere between 28$\%$ and 44$\%$ of human choices. The hybrid confirmation tree is a robust process that can produce accurate hybrid intelligence using existing resources efficiently.

When should organizations consider using the hybrid confirmation tree? We suggest two concrete rules to guide this choice. First, consider moving from using one human to the hybrid confirmation tree. The hybrid confirmation tree might increase costs, as a second human tiebreaker is sometimes necessary to arbitrate human--AI disagreement. However, it reliably reduces errors compared to the average single human (Figure \ref{fig4}). Therefore, as long as the marginal reduction cost due to reduction in errors is greater than the marginal increase in cost due to potentially enlisting one more human opinion, the hybrid confirmation tree is preferable to a single human decision maker. Second, consider moving from using small human-only decision-making structures (e.g., majority vote, hierarchies, or polyarchies) to the hybrid confirmation tree. As long as the AI is as good as the average human decision maker, the hybrid confirmation tree will be as good as the human majority vote and will make error trade-offs between the human-only strategies hierarchy and polyarchy. Empirically, we find that the hybrid confirmation tree is able to improve performance while saving costs (Figures \ref{fig4} and \ref{fig5}). Therefore, as long as the AI is as good as the average human, the hybrid confirmation tree is a cost-effective way to tap into the wisdom of hybrid crowds and to improve on efficiency compared to human-only approaches.

The hybrid confirmation tree is immediately applicable to decision-making problems that typically require multiple independent human decision makers; for example, in medical diagnostics such as mammography screening programs. European Union guidelines recommend using two independent radiologists to diagnose the same mammogram (i.e., double reading) \citep{schunemann2020breast}. Similar consensus mechanisms are used to improve diagnostic accuracies in fields such as breast histopathology \citep{elmore2016evaluation}, skin cancer \citep{tosteson2021association}, liver pathology \citep{hahm2001value}, and ultrasound imaging in search of ovarian cancer \citep{christiansen2025international}. All these examples are centered on the evaluation of images. AI has made significant progress in diagnostic imaging in recent years \citep{gulshan2016development, bejnordi2017diagnostic}, reaching human-level or even superhuman performance \citep{brinker2019deep, brinker2019comparing, rodriguez2019stand}, and the performance gap is expected to widen even further. Medical expertise is costly; relying on multiple humans to finalize a decision should therefore be carefully justified in the face of increases in AI diagnostic accuracy. This trade-off between accuracy and effort is especially pressing in times of ongoing shortages of physicians in general \citep{Darves2017PhysicianShortage, scheffler2019projecting} and in specialties such as radiology \citep{Milburn2024RadiologyWorkforce, Allyn2020IntlRadiologyShortage, RCR2025StateOfTheWait}. Moving from the wisdom of human-only crowds to the wisdom of hybrid crowds using approaches like the hybrid confirmation tree can relieve the burden on medical specialties that rely on image analysis. 

The potential of AI to enhance decision making is widely recognized. Yet although algorithms now achieve remarkable accuracy across diverse fields, the path to widespread AI adoption, especially in high-stakes domains like healthcare and jurisprudence, is bound to be slow. While automating decisions can be appealing, there are several important reasons for constraining the unmediated deployment of AI. People tend to prefer to place trust in human agency, particularly when decisions carry profound personal or ethical consequences \citep{longoni2019resistance, bonnefon2024moral}. Furthermore, legal frameworks such as the European AI Act, professional recommendations \citep{omiye2024large, shahriari2017ieee}, and moral considerations \citep{santoni2021four, habli2020artificial, bonnefon2024moral} all deem accountability and transparency to be essential. To meet this standard, a human must often remain in the loop. Adding to these concerns is the inherent fallibility of AI. Even advanced systems are susceptible to errors, biases, and unexpected failures \citep{mcgregor2021preventing, angwin2016machine, wong2021external, omiye2023large, shojaee2025illusion}. The challenges of trust, accountability, and reliability restrict the direct application of AI where it might otherwise offer substantial benefits. This is where the hybrid confirmation tree---which embeds human oversight and ensures that every final decision is explicitly approved by at least one human---offers a promising way forward. The sequential process, consisting of independent initial assessments by a human and an AI, followed by a human tiereaker in cases of disagreement, inherently guarantees human agency and accountability. This architecture may therefore unlock opportunities to leverage AI in sensitive areas where full automation is currently untenable, offering a pragmatic and responsible method for combining the competencies of humans and machines.

The hybrid confirmation tree is a simple and robust strategy to combine human and AI choices. Our theoretical analysis and empirical evaluation across six datasets show valuable performance increases over human-only decision making strategies at reduced costs. We expect similar results in other domains---possibly in multiple choice and open-ended problems as well. The hybrid confirmation tree can improve decision making in the real world by efficiently tapping into the wisdom of hybrid crowds.






\section*{Methods}

\subsection*{Data and code availability}

All data wrangling, analyses, and plotting was done using R 4.4.2. \citep{r}. Code and data used to generate results and figures can be accessed at \url{https://osf.io/ytbmx/?view_only=c727e09949c64a0c94a4faef77cdeda1}.

\subsection*{Modeling the hybrid confirmation tree}

To formally compare the hybrid confirmation tree with a three-human majority vote, we derived analytical expressions for their accuracy and expected cost (i.e., the number of human decision makers consulted). We assumed binary decisions where each agent (human or machine) makes an independent judgment and has a specific probability of being correct. Let:
\begin{itemize}
    \item $p_H$ be the probability that a human makes a correct decision (human accuracy), and
    \item $p_M$ be the probability that the machine makes a correct decision (machine accuracy).
\end{itemize}
We assumed all humans have the same accuracy $p_H$, and all decisions are statistically independent.

The hybrid confirmation tree procedure is as follows (Figure \ref{tree}):
\begin{enumerate}
    \item Human 1 (H1) and the machine (M) make independent decisions.
    \item If H1 and M agree, their consensus decision is the final outcome.
    \item If H1 and M disagree, a second human (H2) makes a decision, which becomes the final outcome.
\end{enumerate}

The hybrid confirmation tree makes a correct final decision in three mutually exclusive scenarios:
\begin{itemize}
    \item Scenario 1: H1 is correct, M is correct. Probability: $p_H p_M$.
    \item Scenario 2: H1 is correct, M is incorrect, H2 is correct. Probability: $p_H (1-p_M) p_H$.
    \item Scenario 3: H1 is incorrect, M is correct, H2 is correct. Probability: $(1-p_H) p_M p_H$.
\end{itemize}
The total accuracy of the hybrid confirmation tree is the sum of these probabilities:
\begin{align}
    \pi_{HCT} &= P(\text{H1 correct, M correct}) \nonumber \\
              &\quad + P(\text{H1 correct, M incorrect, H2 correct}) \nonumber \\
              &\quad + P(\text{H1 incorrect, M correct, H2 correct}) \nonumber \\
    \pi_{HCT} &= p_H p_M + p_H (1-p_M) p_H + (1-p_H) p_M p_H \nonumber .
\end{align}

H1 is always involved, contributing 1 to the cost. H2 is consulted only if H1 and M differ in their initial decisions. Thus, the expected cost can be expressed as:
\begin{equation*}
    C_{HCT} = 1 + P(\text{H1 and M disagree}) . \nonumber
\end{equation*}
The probability that H1 and M disagree is $1 - P(\text{H1 and M agree})$. Agreement occurs if both are correct---probability $p_H p_M$---or if both are incorrect---probability $(1-p_H)(1-p_M)$. Therefore:
\begin{align*}
    P(\text{H1 and M disagree}) &= \\
    1 - [p_H p_M + (1-p_H)(1-p_M)] .
\end{align*}
Substituting this into $C_{HCT} = 1 + P(\text{H1 and M disagree})$ results in:
\begin{align}
    C_{HCT} &= 1 + \{1 - [p_H p_M + (1-p_H)(1-p_M)]\} \nonumber \\
            &= 2 - [p_H p_M + (1-p_H)(1-p_M)] \label{eq:hct_cost_meth_final_alt_meth} .
\end{align}
This shows that the cost is 2 minus the probability that H1 and M independently agree.
\\
\\
The human majority vote works similarly. If H1 and H2 agree, their decision is accepted. If they disagree, H3 breaks the tie. The majority vote is correct if:
\begin{itemize}
    \item Scenario 1: H1 correct, H2 correct. Probability: $p_H^2$.
    \item Scenario 2: H1 correct, H2 incorrect, H3 correct. Probability: $p_H (1-p_H) p_H$.
    \item Scenario 3: H1 incorrect, H2 correct, H3 correct. Probability: $(1-p_H) p_H p_H$.
\end{itemize}

The total accuracy of the majority vote is:
\begin{align}
    \pi_{MAJ} &= P(\text{H1 correct, H2 correct}) \nonumber \\
              &\quad + P(\text{H1 correct, H2 not, H3 correct}) \nonumber \\
              &\quad + P(\text{H1 not, H2 correct, H3 correct}) \nonumber \\
    \pi_{MAJ} &= p_H^2 + p_H (1-p_H) p_H + (1-p_H) p_H p_H \nonumber \label{eq:maj_accuracy_meth} .
\end{align}

To calculate the cost of the majority vote we have to consider that H1 and H2 are always consulted and H3 is only consulted if H1 and H2 disagree.
The probability that H1 and H2 agree is $P(\text{H1,H2 agree}) = p_H^2 + (1-p_H)^2$.
The expected cost $C_{MAJ}$ is:
\begin{align}
    C_{MAJ} &= 2 \cdot P(\text{H1,H2 agree}) + \\& 3 \cdot P(\text{H1,H2 disagree}) \nonumber \\
            &= 2 \cdot [p_H^2 + (1-p_H)^2] + \\ & 3 \cdot [1 - (p_H^2 + (1-p_H)^2)] \nonumber \\
            &= 3 - [p_H^2 + (1-p_H)^2] . \nonumber 
\end{align}

\subsection*{Modelling correlated decisions}   

To investigate the interplay between human and machine accuracy, as well as the dependence between human and machine decisions, we used an opinion-leader model, where one follower agrees with a leader's binary choice with a given probability \citep{grofman1983thirteen, kurvers2019detect}. That is, we designated one simulated human decision maker to be the leader, enabling us to vary the likelihood $\alpha$ of both the humans and the machine that they agreed with the human decision leader. When $\alpha = 1$, the follower always copies the leader, yielding perfect correlation between their decisions. When $\alpha = 0$, decisions are independent, and any agreement arises from their individual accuracies. Intermediate values of $\alpha$ create partial dependence, where the follower sometimes relies on the leader and sometimes decides independently. This setup allowed us to simulate the hybrid confirmation tree with various human--machine and human--human choice dependencies, as well as to investigate how the human majority vote was impacted by human--human choice dependence.

Increasing accuracies of humans and machines inevitably lead to high agreement on choices between agents. We therefore rely on Cohen's $\kappa$ to quantify how related the decisions between two agents (human or machine) were. Here $\kappa$ was expressed as a value between $-1$ and 1, where 0 represented chance agreement. 

The general data generation mechanism integrated both the opinion-leader model \citep{grofman1983thirteen} and $\kappa$ \citep{cohen1960coefficient}. The data generation posited that a follower agent ($A_F$) determines its binary labels based on a leader agent ($A_L$) as follows: With a copy probability $\alpha$, the follower adopts the leader's label: $R_{Fi} = R_{Li}$. With probability $1-\alpha$, the follower disregards the leader and generates an independent label from a Bernoulli distribution with success likelihood $p'$ (the independent draw rate of the follower).
Let $p_L = P(A_L=1)$ denote the leader's marginal success probability and $p_F = P(A_F=1)$ denote the follower's marginal success probability.

Our objective was that the dependence between a leader and a follower yields a prespecified Cohen's $\kappa$ and ensures the follower achieves a prespecified marginal accuracy $p_F$. The analytical relationship between these constructs is as follows:

\begin{itemize}
    \item \textbf{Follower's marginal accuracy ($p_F$):} The follower's overall success probability is a weighted average of its success probability in a mixture model that consists of the leader's accuracy $p_L$, the likelihood to copy the leader $\alpha$, and the follower's independent draw rate $p'$:
    
    \begin{align*}
        p_F &= \alpha P(A_F=1 | \text{copy}) + (1-\alpha) P(A_F=1 | \text{indep. draw}) \\
        p_F &= \alpha p_L + (1-\alpha)p' .
    \end{align*}

    \item \textbf{Observed agreement ($p_o$):} Cohen's $\kappa$ is defined as 
    $$\kappa = \frac{p_o - p_e}{1 - p_e},$$ 
    where $p_o$ is the rate of observed agreement between follower and leader $P(A_L=A_F)$. We calculated it as:
    
    \begin{align*}
        p_o &= \alpha P(A_L=A_F | \text{copy}) + (1-\alpha) P(A_L=A_F | \text{independent draw}) \\
        p_o &= \alpha \cdot 1 + (1-\alpha) [p_L p' + (1-p_L)(1-p')] .
    \end{align*}    

    \item \textbf{Chance agreement ($p_e$):} For Cohen's $\kappa$, chance agreement was calculated based on the observed marginal probabilities of the two decision makers:
    \begin{equation*}
        p_e = p_L p_F + (1-p_L)(1-p_F).
    \end{equation*}
    
    \item \textbf{Agreement beyond chance ($p_o - p_e$):} Intuitively, the excess agreement beyond what one would expect by chance is directly proportional to the copy probability $\alpha$. The more often the follower copies the leader, the more their agreement exceeds chance levels. Specifically,
    \begin{equation*}
        p_o - p_e = 2\alpha p_L (1-p_L),
    \end{equation*}

since
\[
\begin{aligned}
p_o - p_e
&= \bigl[\alpha + (1-\alpha)(p_Lp' + (1-p_L)(1-p'))\bigr]\\
&\quad
  -\bigl[\alpha(p_L^2+(1-p_L)^2)+\\&(1-\alpha)(p_Lp' + (1-p_L)(1-p'))\bigr]\\
&= \alpha - \alpha\bigl[p_L^2 + (1-p_L)^2\bigr]
  \\
&= \alpha\bigl[1 - (p_L^2 + (1-p_L)^2)\bigr]
\end{aligned}
\]
and 
\[
1 - \bigl[p_L^2 + (1-p_L)^2\bigr]
= 2\,p_L\,(1-p_L).
\]
    
    This shows that the excess agreement is the product of three factors: the copying rate ($\alpha$), the leader's accuracy ($p_L$), and the leader's error rate ($1-p_L$). The term $2p_L(1-p_L)$ is maximized when $p_L = 0.5$, meaning dependence has the strongest effect when the leader's accuracy is at chance level.
\end{itemize}

While the above equations express $\kappa$ and $p_F$ as functions of $\alpha$ and $p'$, our investigation required the reverse; namely, determining the $\alpha$ and $p'$ that will produce a desired $\kappa$ and a desired $p_F$.

\begin{itemize}
    \item \textbf{Solving for copy probability ($\alpha$):}
    We set the desired follower accuracy $p_F$. The denominator of $\kappa$, $1-p_e$, becomes $1 - [p_L p_F + (1-p_L)(1-p_F)]$, which simplifies to $p_L(1-p_F) + (1-p_L)p_F$.
    Using $p_o - p_e = 2\alpha p_L (1-p_L)$ for the numerator, Cohen's $\kappa$ can be expressed as:
    $$ \kappa = \frac{2\alpha p_L (1-p_L)}{p_L(1-p_F) + (1-p_L)p_F} .$$
    Given a target $\kappa$, we solve for $\alpha$:
    \begin{equation*}
        \alpha = \kappa \frac{p_L(1-p_F) + (1-p_L)p_F}{2 p_L (1-p_L)} .
    \end{equation*}
    
    This equation provides the necessary copy probability $\alpha$ to achieve the target $\kappa$ when the leader's accuracy is $p_L$ and the follower's target accuracy is $p_F$. This calculation can be performed for each leader--follower pair (i.e., both human--human and human--machine) using their respective target $\kappa$ and accuracies.

    \item \textbf{Solving for independent draw rate of the follower ($p'$):}
    Once $\alpha$ was determined using the target $\kappa$ and $p_F$, we used the follower's marginal accuracy to solve for the independent draw rate $p'$:
    $$ p_F = \alpha p_L + (1-\alpha)p' $$
    $$ p' = \frac{p_F - \alpha p_L}{1-\alpha}.$$
    This $p'$ ensures that the follower's marginal accuracy matches the target $p_F$, given the calculated $\alpha$ and the leader's $p_L$.
\end{itemize}

The necessity of solving for $\alpha$ and $p'$ highlights that not all combinations of $(p_L, p_F, \kappa)$ are feasible. The calculated copy probability $\alpha$ from $\alpha = \kappa \frac{p_L(1-p_F) + (1-p_L)p_F}{2 p_L (1-p_L)}$ must be within the range $[0,1]$. If $\kappa$ is too high relative to the agreement potential allowed by the disparity between $p_L$ and $p_F$, the equation can yield $\alpha > 1$ or $\alpha < 0$. For instance, if $p_L$ and $p_F$ are widely different (e.g., $p_L=0.9, p_F=0.1$), the term $p_L(1-p_F) + (1-p_L)p_F$ (which is $1-p_e$, the denominator of $\kappa$) can be large. If $2p_L(1-p_L)$ is small (as it is when $p_L$ is near 0 or 1), the ratio $\frac{p_L(1-p_F) + (1-p_L)p_F}{2 p_L (1-p_L)}$ can be very large. Multiplying this by even a moderate $\kappa$ might result in $\alpha > 1$. This implies that the desired level of dependence between leader and follower decisions cannot be achieved through the copy mechanism under the given accuracy constraints. Similarly, the calculated independent draw rate $p'$ from must also be within $[0,1]$. Even if $\alpha$ is valid, an extreme $p_F$ relative to $\alpha p_L$ might force $p'$ outside its permissible range. For example, if $p_F$ is significantly higher than $\alpha p_L$ (the accuracy achieved by copying), and $1-\alpha$ is small (frequent copying), the term $(p_F - \alpha p_L)/(1-\alpha)$ can exceed 1, indicating that the independent draws cannot sufficiently compensate to reach $p_F$.

\subsubsection*{Hybrid confirmation tree with correlated decisions.}

To derive an analytical solution for the hybrid confirmation tree that accounts for decision dependence, we extended our notation. 

The relationship between the leader and each follower is parameterized by a target Cohen's $\kappa$, which determines a copy probability ($\alpha$) and an independent draw rate ($p'$). The H1--M relationship is defined by $\kappa_{HM}$, yielding $\alpha_{HM}$ and $p'_{HM}$. The H1--H2 relationship is defined by $\kappa_{HH}$, yielding $\alpha_{HH}$ and $p'_{HH}$.

The core of the accuracy derivation is the conditional probability of a follower being correct, given the leader's outcome. For any follower $F$ with parameters $(\alpha_F, p'_F)$ and leader $L$:
\begin{itemize}
    \item Given leader is correct: The follower is correct if it copies the correct decision (probability $\alpha_F$) or makes an independent correct decision (probability $(1-\alpha_F)p'_F$),
    \begin{equation*}
        P(\text{F correct} \mid \text{L correct}) = \alpha_F + (1-\alpha_F)p'_F .
    \end{equation*}
    
    \item Given leader is incorrect: The follower is correct only if it does not copy the leader's incorrect decision and instead makes a correct independent draw,
    \begin{equation*}
        P(\text{F correct} \mid \text{L incorrect}) = (1-\alpha_F)p'_F .
    \end{equation*}
\end{itemize}

The overall accuracy of the hybrid confirmation tree, $\pi_{HCT}$, is the sum of probabilities of three mutually exclusive scenarios where a correct final decision is reached. Since the followers (M and H2) are conditionally independent given the leader's (H1) decision, the joint probabilities for the three success scenarios can be expressed as follows:
\begin{itemize}
    \item \textbf{Scenario 1: H1 and M are correct.} They agree, and the process stops with a correct outcome:
    \begin{align*}
        P(\text{H1 correct, M correct}) &= P(\text{M correct}\mid\text{H1 correct}) \cdot P(\text{H1 correct}) \\
                       &= \left[ \alpha_{HM} + (1-\alpha_{HM})p'_{HM} \right] \cdot p_{H1} .
    \end{align*}
    
    \item \textbf{Scenario 2: H1 and H2 are correct but M is not.} H1 and M disagree, so H2 is consulted and makes the correct tiebreaking decision:
\begin{align*}
&P(\text{H1 correct, M incorrect, H2 correct}) \\
&= P(\text{M incorrect}\mid\text{H1 correct}) \cdot P(\text{H2 correct}\mid\text{H1 correct}) \cdot P(\text{H1 correct}) \\
&= \left( 1 - [\alpha_{HM} + (1-\alpha_{HM})p'_{HM}] \right) \\
&\quad \cdot [\alpha_{HH} + (1-\alpha_{HH})p'_{HH}] \cdot p_{H1} .
\end{align*}
    
    \item \textbf{Scenario 3: The H1 is incorrect but M and H2 are correct.} H1 and M disagree, so H2 is consulted and makes the correct tiebreaking decision:
\begin{align*}
&P(\text{H1 incorrect, M correct, H2 correct}) \\
&= P(\text{M correct}\mid\text{H1 incorrect}) \cdot P(\text{H2 correct}\mid\text{H1 incorrect}) \cdot P(\text{H1 incorrect}) \\
&= [(1-\alpha_{HM})p'_{HM}] \cdot [(1-\alpha_{HH})p'_{HH}] \cdot (1-p_{H1}) .
\end{align*}

\end{itemize}

The total accuracy of the hybrid confirmation tree with correlated decisions, $\pi_{\text{HCT, cor}}$, is the sum of the probabilities from the above scenarios:
\begin{align} 
\pi_{\text{HCT, cor}} &= p_{H1} \left[ \alpha_{HM} + (1-\alpha_{HM})p'_{HM} \right] \nonumber \\
&\quad + p_{H1} \left( 1 - [\alpha_{HM} + (1-\alpha_{HM})p'_{HM}] \right) \nonumber \\
&\quad \cdot \left( \alpha_{HH} + (1-\alpha_{HH})p'_{HH} \right) \nonumber \\
&\quad + (1-p_{H1}) (1-\alpha_{HM})p'_{HM} \nonumber \\
&\quad \cdot (1-\alpha_{HH})p'_{HH}.
\end{align}

\subsubsection*{Majority vote with correlated decisions.}

To derive an analytical solution for the majority vote (MAJ) accuracy that accounts for decision dependence, we assumed all humans had the same target accuracy ($p_{H1}=p_{H2}=p_{H3}=p_H$) and the same dependency on the leader ($\kappa_{HH}$, $\alpha_{HH}$, $p'_{HH}$). Generally, the process was similar to the one outlined for the hybrid confirmation tree. For brevity, let $P(\text{H follower correct}\mid\text{H1 correct}) = \alpha_{HH} + (1-\alpha_{HH})p'_{HH}$ be the probability of a human follower being correct given H1 is correct, and let $P(\text{H follower correct}\mid\text{H1 incorrect}) = (1-\alpha_{HH})p'_{HH}$ be the probability of a human follower being correct given H1 is incorrect.

A correct final decision is reached in one of three mutually exclusive ways.

\begin{itemize}
    \item \textbf{Scenario 1: H1 and H2 are both correct.} They agree, and the process stops with a correct outcome:
    \begin{align*}
        P(\text{H1 correct, H2 correct}) &= P(\text{H2 correct}\mid\text{H1 correct}) \cdot P(\text{H1 correct}) \\
        &= P(\text{H follower correct}\mid\text{H1 correct}) \cdot p_H .
    \end{align*}
    
    \item \textbf{Scenario 2: H1 is correct, H2 is incorrect, and H3 is correct.} H1 and H2 disagree, so H3 is consulted and makes the correct tiebreaking decision:
    \begin{align*}
        &P(\text{H1 correct, H2 incorrect, H3 correct}) \\
        &= P(\text{H2 incorrect}\mid\text{H1 correct}) \cdot P(\text{H3 correct}\mid\text{H1 correct}) \cdot P(\text{H1 correct}) \\
        &= \left[1-P(\text{H follower correct}\mid\text{H1 correct})\right] \cdot P(\text{H follower correct}\mid\text{H1 correct}) \cdot p_H .
    \end{align*}
    
    \item \textbf{Scenario 3: H1 is incorrect, H2 is correct, and H3 is correct.} H1 and H2 disagree, so H3 is consulted and makes the correct tiebreaking decision:
    \begin{align*}
        &P(\text{H1 incorrect, H2 correct, H3 correct}) \\
        &= P(\text{H2 correct}\mid\text{H1 incorrect}) \cdot P(\text{H3 correct}\mid\text{H1 incorrect}) \cdot P(\text{H1 incorrect}) \\
        &= \left[P(\text{H follower correct}\mid\text{H1 incorrect})\right]^2 \cdot (1-p_H) .
    \end{align*}
\end{itemize}

The total accuracy, $\pi_{\text{MAJ, cor}}$, is the sum of these three scenario probabilities:
\begin{align} \label{eq:maj_correlated_accuracy}
\pi_{\text{MAJ, cor}} &= p_H \left( \alpha_{HH} + (1-\alpha_{HH})p'_{HH} \right) \nonumber \\
&\quad + p_H \left( 1 - [\alpha_{HH} + (1-\alpha_{HH})p'_{HH}] \right) \nonumber \\
&\quad \cdot \left( \alpha_{HH} + (1-\alpha_{HH})p'_{HH} \right) \nonumber \\
&\quad + (1-p_H) [(1-\alpha_{HH})p'_{HH}]^2.
\end{align}

\subsection*{Bayesian estimation}

We employed Bayesian estimation to quantify the accuracy difference between the hybrid conformation tree and the human majority vote. We relied on the general framework by \citet{benavoli2017time} to account for correlated samples (i.e., accuracy was calculated on the same cases within datasets) and quantified the difference using a region of practical equivalence (ROPE) approach of $\pm$ 1 percentage point. We implemented this framework using \texttt{brms} \citep{burkner2017brms} in R 4.4.2 \citep{rcore}. To test the general difference, we modeled the accuracy of the hybrid confirmation tree and the human majority vote via a generalized linear mixed regression model (GLMM) with no-intercept an a hierarchical structure of cases contained within datasets using the formula \texttt{accuracy $\sim$ 0 + strategy + (1 | case)}, where strategy is a factor indicating either the hybrid confirmation tree or the human majority vote and accuracy is the accuracy either method achieved on a case. We ran all GLMMs with four chains and 10,000 iterations after warm-up and checked chain convergence using the Gelman--Rubin criterion $\hat{R} <$ 1.01. Using the fitted GLMMs, we employed \texttt{marginaleffects} \citep{arel2024interpret} to compute the difference and 95$\%$ highest density interval (HDI) between the conditional marginal posterior medians of the accuracy distributions of the hybrid confirmation tree and the human majority vote. We used \texttt{bayestestR} \citep{makowski2019bayestestr} to compute the probability of direction (percentage of the posterior difference that is greater than 0) and the practical significance (percentage of the posterior difference that is greater than the top end of the ROPE, i.e., 1 percentage point).

\subsection*{Cross-validation}

We cross-validated the results presented in Figure \ref{fig4} by testing whether out-of-sample selection of the accuracy-maximizing threshold for the model contained in the hybrid confirmation tree returned the same results. We used a 1,000-times repeated 5-fold cross-validation over cases within a dataset from resamples. The simulation procedure described in the main text (Improving decision making in high-stakes domains) provided us with all hybrid confirmation tree results of all possible human--AI triplets for every threshold setting possible over all cases. Our cross-validation resampled from these results in the following way: Per dataset, we split the cases into five equal-sized folds over all cases in a stratified manner, thereby maintaining the base rate of positive and negative cases of the whole data in the folds. Next, we took four of the five folds as a train set and calculated the threshold setting of the machine at which the hybrid confirmation tree achieved the maximum accuracy. We then selected the one remaining fold of the five---the test set---and filtered for the threshold that maximized the accuracy in the train set and calculated the accuracy of the hybrid confirmation tree as an average over all cases. For every train and test set, we also noted the accuracy of the majority vote, individual humans, and the machine alone. This procedure was repeated 1,000 times. 

Finally, we again used Bayesian estimation and estimated the accuracy of the four methods (i.e., human individual, human majority vote, machine alone, and hybrid confirmation tree) based on these resamples. As suggested by \citet{benavoli2017time}, we modeled the accuracy of the results taking into account the dependence that results were generated by folds nested in repetitions; our formula was therefore \texttt{accuracy $\sim$ 0 + strategy + (1 | repetition/fold)}. Model specifications remained the same as described in the section \textit{Bayesian estimation}. Finally, we calculated the marginal contrast between means of the estimated accuracy posteriors; the results are presented in Supplementary Figure \ref{si_cv_marginals}.

\subsection*{Estimating Cohen's $\kappa$ from data}

We estimated human--human ($\kappa_{HH}$) and human--machine ($\kappa_{HM}$) correlations from data on a individual level basis. To calculate $\kappa_{HH}$ for one target human, we referenced all other humans who made choices with the target human on the same cases. We then calculated $\kappa_{HH}$ using \texttt{irr::kappa2()} \citep{irr} for every pair the target human belonged to and averaged across all pairs for the target human. To calculate $\kappa_{HM}$ we used the same R-function among all cases that a human and the machine shared. We repeated these two steps for all humans.

\subsection*{Data}\label{sec:Data}

\subsubsection*{Dermatology.}\label{sec:Dermoscopy}

We used the Melanoma Classification Benchmark provided by the German Cancer Research Center, described in \citet{brinker2019deep} and \citet{brinker2019comparing}.\footnote{Although the public download link to the data is no longer available, the website was preserved by the Internet Archive at \url{https://web.archive.org/web/20210620070355/https://skinclass.de/mclass/.}} The benchmark consists of two datasets: one of dermoscopic images (i.e., taken with a dermoscope) and one of nondermoscopic images (i.e., taken with any other camera). Each dataset consists of 100 images of skin lesions, of which 20 are melanoma. The binary ground truth of these skin lesions (i.e., benign lesion versus melanoma) is based on histopathology. The dermoscopic images were diagnosed by 157 dermatologists (resulting in 15,700 human choices) and the nondermoscopic images by 145 dermatologists (resulting in 14,500 choices). Dermatologists classified images as either benign or melanoma.

We developed a convolutional neural network (CNN) to classify the images. We trained our CNN on more than 58,000 images taken from publicly available datasets from the International Skin Imaging Collaboration (ISIC) challenges of 2019 and 2020 \citep{tschandl2018ham10000, rotemberg2021patient, codella2018skin, hernandez2024bcn20000}. The training procedure is described in \citet{andersen2023confirmation}. Before training, we ensured that the images we tested were not also part of the data we used for training, as the original authors of the Melanoma Classification Benchmark also used images from publicly available sources. We thereby avoided leakage and our CNN's performance was not tainted by having already seen the images it was tested on. Applying our CNN to the images in both datasets resulted in area under the curve (AUC) values of 0.89 (dermoscopic images) and 0.90 (nondermoscopic images).

\subsubsection*{Deepfakes.}\label{sec:Deepfakes}

For the deepfakes data, we used the data made available by \citet{groh2022deepfake}, which consists of 54 videos from the Deepfake Detection Challenge. Of these, half were deepfakes \citep{dolhansky2020deepfake}. We focused our analyses on the data collected by Groh et al. in Experiment 2: Single Video Design, in which participants recruited on Prolific indicated whether a single video was a deepfake on a slider ranging from ``100\% confidence this is not a deepfake'' to ``100\% confidence this is a deepfake'' (midpoint: ``just as likely a deepfake as not a deepfake''). Participants' judgments at the midpoint were randomly assigned as either a deepfake or a no-deepfake choice. We restricted our analysis to participants in the control conditions who passed the attention check, resulting in 1,347 deepfake classifications made by 132 individuals on 54 videos. 

As AI predictions we used the predictions of the winning model in the Deepfake Detection Challenge by \citet{seferbekov2021deepfake}. This model achieved an AUC of 0.95.

\subsubsection*{Criminal rearrest.}\label{sec:Criminal re-arrest}

For predicting criminal rearrest, we used the COMPAS dataset \citep{angwin2016machine}, featuring open data collected by \citet{dressel2018accuracy}. Binary human predictions of criminal rearrest within 2 years were collected on Amazon Mechanical Turk. Dressel and Farid had participants evaluate two sets of case files---one that included racial information on the defendant (COMPAS [race]) and one that did not (COMPAS [no race]). We focused our analysis on the case files without racial information. The data consisted of 1,000 case files with a criminal rearrest prevalence of 47$\%$. A total of 400 participants evaluated 50 cases each, resulting in 20,000 human evaluations.

We developed a small logistic regression model to predict the likelihood of criminal rearrest. Sparse models (i.e., models using only a few features) performed well on the COMPAS data \citep{dressel2018accuracy, jung2020simple}, so we used a logistic regression model for simplicity's sake. The model had access to the features \textit{age} and \textit{count of prior nonjuvenile crimes}. The COMPAS data consists of 7,214 cases in total, of which 1,000 were used as the test set by \citet{dressel2018accuracy}. We used the remaining 6,214 cases as the training set for our logistic model. The resulting model had an AUC of 0.72 on the test set.

\subsubsection*{Geopolitical forecasting.}

We studied two datasets on geopolitical forecasts. First, we evaluated openly accessible data from the Hybrid Forecasting Competition (HFC) \citep{benjamin2023hybrid}. The HFC aimed to compare how humans and time-series forecasting models make predictions about geopolitical events and whether humans with access to model predictions could improve their forecasts. Forecasts had either multiple possible outcomes or just two (i.e., event will vs. will not happen by a given point in time); we restricted our analyses to events with just two outcomes. Participants forecasted the likelihood that an event would occur by a resolution date on a probabilistic scale between 0\% and 100\%. We converted all probabilistic forecasts below 50\% to the choice of 0 (event will not happen) and all forecasts above 50\% to 1 (event will happen); we randomly assigned either 0 or 1 to exact 50\% forecasts. We restricted our analyses to human participants who provided more than five estimates in the experimental condition B of \citet{benjamin2023hybrid}, in which participants saw time series information and newspaper articles for a forecasting question if available but did not have access to any model predictions. All participants were recruited on Amazon Mechanical Turk. For the AI, we chose the ensemble model PHE2, since it performed best in the competition. During the HFC, human participants and the models could update their forecasts over time. As the certainty of an event happening or not is likely to increase as time nears the resolution date, we restricted the selection of our human and model forecasts to those made at least 7 days before the resolution date for every forecasting question. For example, for a human who forecasted an event's probability 10 days before the resolution date and updated their belief the day before resolution, we would have disregarded the later forecast and used the 10-day forecast. Our approach was the same for the AI forecasts. Mirroring previous work \citep{satopaa2021bias}, we took this approach in order to ensure some uncertainty remained in the forecasting problem. Altogether, we had 1,055 forecasts made by 111 humans on 52 forecasting questions.

Second, we analyzed data made available on ForecastBench, an ongoing evaluation of human and large language model (LLM) capabilities in forecasting geopolitical questions \citep{karger2025forecastbench}. ForecastBench collects questions from forecasting tournaments and prediction markets and periodically allows humans and teams that operate LLMs to update forecasts over time. We restricted our analyses to questions that pertained to one event (e.g., ``What is the likelihood that event A happens?") and that had been resolved by January 17, 2025. We discarded all compound questions (e.g., ``What is the likelihood that event A and event B happen?"). Human participants were recruited on Prolific and Facebook. We binarized human probability forecasts in the same manner as for the HFC data. Although ForecastBench provides a suite of models to compare against each other, we restricted our analyses to the forecasts by ``Claude-3-5-Sonnet-20240620 (scratchpad with freeze values)," which was the leading LLM as of March 17, 2025. Human participants had 7 days to participate until forecasts were due, whereas LLMs received the forecasts 24 hours before the due date so as to avoid manual human steering the LLM forecasting ability due to lack of time. This process allowed us to analyze 21,302 forecasts by 500 human participants on 422 cases.

\section*{Author contributions – CRediT statement}

Conceptualization: J.B., P.P.A., F.A., V.S., and R.H.J.M.K. Data curation: J.B. and V.S. Formal analysis: J.B., P.P.A., V.S., and R.H.K. Funding acquisition: J.B., P.P.A., and R.H.K. Investigation: J.B., P.P.A., V.S., and R.H.K. Methodology: J.B., P.P.A., F.A., V.S., and R.H.J.M.K. Project administration: J.B., P.P.A., and R.H.K. Resources: J.B., P.P.A., K.P.L., V.S., and R.H.K. Software: J.B., K.P.L., and V.S. Supervision: P.P.A., V.S., and R.H.K. Validation: J.B., P.P.A., V.S., and R.H.K. Visualization: J.B., P.P.A., F.A., V.S., and R.H.K. Writing - original draft: J.B., P.P.A., V.S., and R.H.K. Writing - review $\&$ editing: J.B., P.P.A., F.A., K.P.L., V.S., and R.H.K.

\bibliographystyle{plainnat} 
\bibliography{ref_opt} 



\newpage
\appendix 

\setcounter{figure}{0}
\setcounter{table}{0}

\begin{figure}[!htbp]
    \centering
    \includegraphics[width = \textwidth]{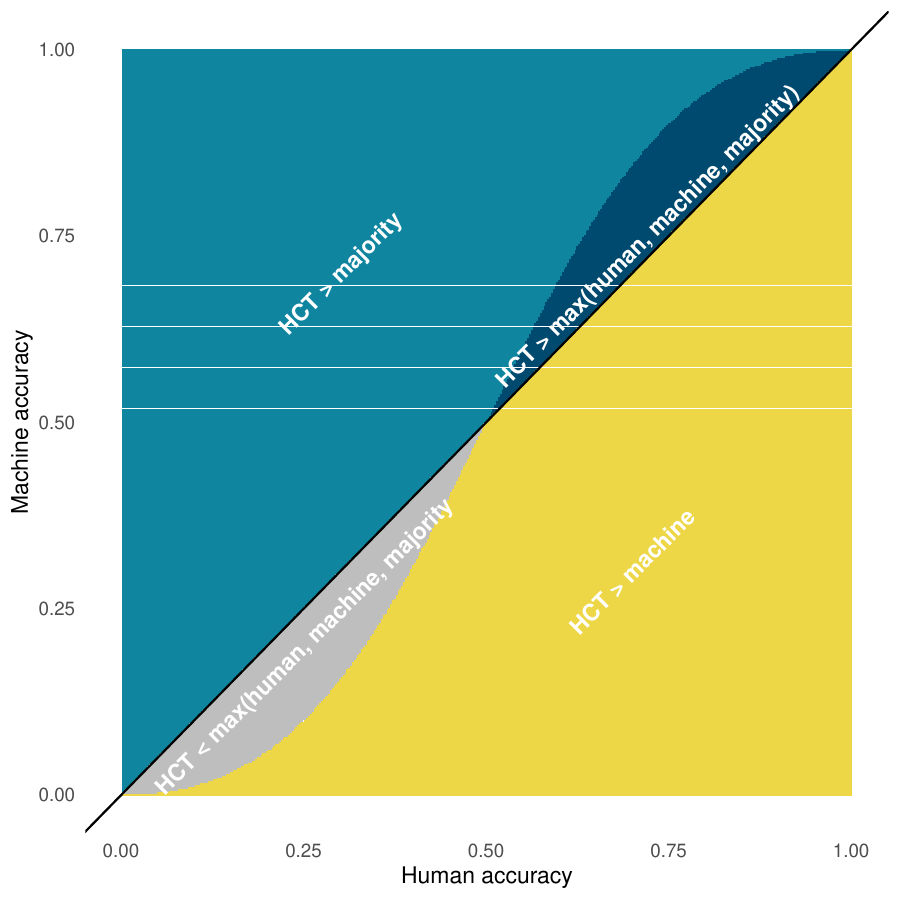}
    \caption{Comparison of the accuracy of the hybrid confirmation tree, single humans, a machine, and a three-person majority vote as a function of human accuracy and machine accuracy. Regions where the hybrid confirmation tree outperforms only the majority vote (blue), only the machine (yellow), and both the majority vote and the machine (dark blue), as well as a region where the hybrid confirmation tree performs worse than the majority vote and the machine alone (gray). }
    \label{si_full_theory}
\end{figure}

\begin{figure}[!htbp]
    \centering
    \includegraphics[width = \textwidth]{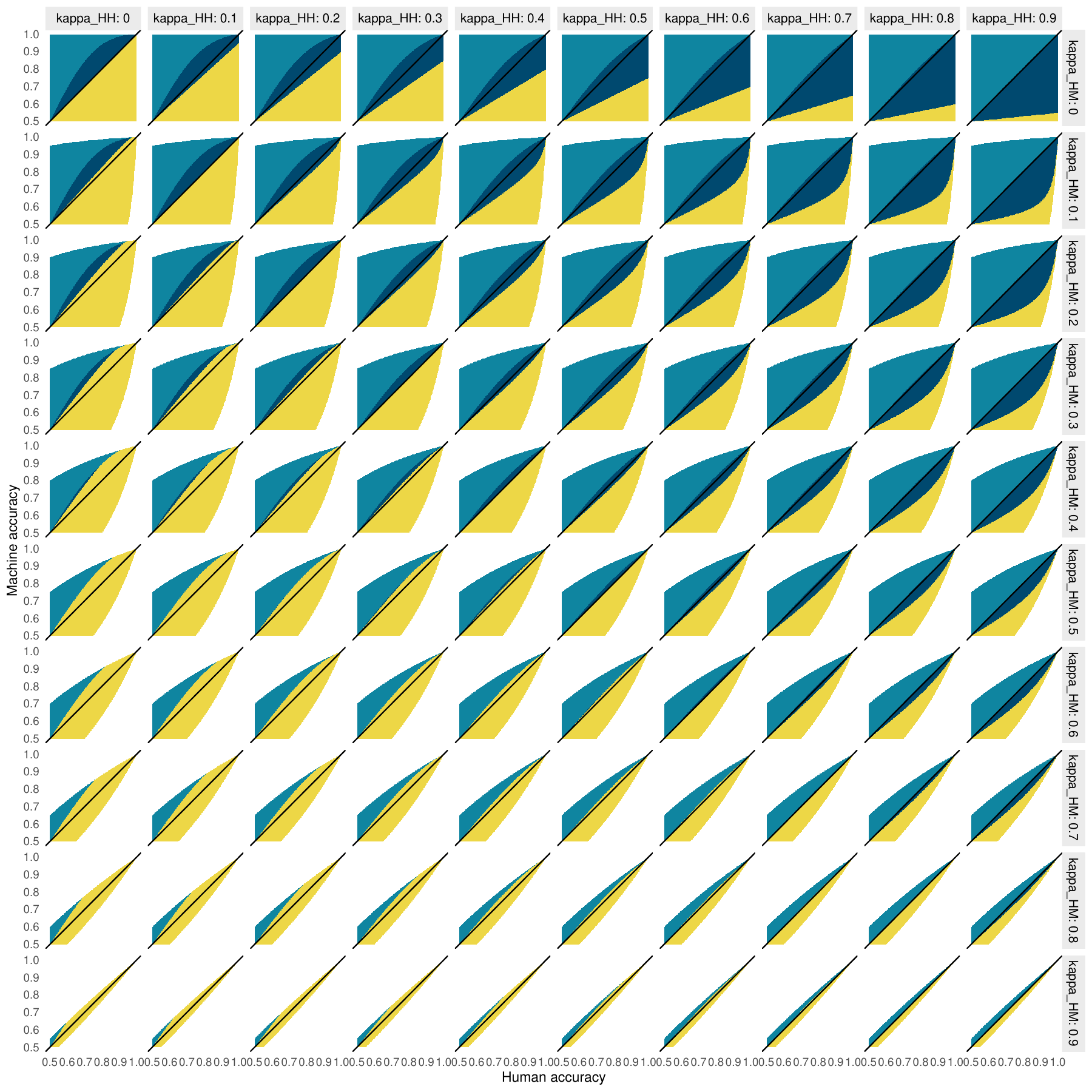}
    \caption{Comparison of the hybrid confirmation tree, single humans, a machine, and a three-person majority vote as a function of human accuracy and machine accuracy. Colors indicate regions where the hybrid confirmation tree outperforms the human majority vote (blue), the machine (yellow), and both the human majority vote and the machine (dark blue), and where it performs worse than the machine and the human majority vote (gray). Columns and rows show increasing human--human ($\kappa_{HH}$) and human--machine ($\kappa_{HM}$) decision making correlation. Higher $\kappa$ values are only possible for more similar human and machine accuracies (i.e., large differences between human and machine accuracies result in lower $\kappa$ values).}
    \label{si_kappa}
\end{figure}

\newpage

\begin{table}[ht]
\centering
\caption{Posterior estimates of accuracy difference between the hybrid confirmation tree and the human majority vote. Estimates are marginal median difference between the posterior estimates of hybrid confirmation tree accuracy and majority vote accuracy; upper and lower correspond to the 95$\%$ highest density intervals of that estimate. PS = probability of significance, a measure that quantifies how much of the posterior difference is greater than 1 percentage point.}
\label{tab:posterior}
\begin{tabular}{@{}lrrrrr@{}}
\toprule
Domain              & Estimate &   Lower    &   Upper   &   PS  & $\hat{R}$ \\ 
\midrule
Skin Cancer (Derm)         & -0.1035 & -0.0786 & -0.1283 & 1.00  & 1.00 \\
Skin Cancer (Nonderm)     & -0.0854 & -0.0564 & -0.1132 & 1.00  & 1.00 \\
Deepfakes           & -0.0790 & -0.0348 & -0.1260 & 0.998 & 1.00 \\
Criminal Rearrest    & -0.0233 & -0.0141 & -0.0322 & 0.997 & 1.00 \\
Hybrid Forecasting Comp                 & -0.1037 & -0.0840 & -0.1230 & 1.00  & 1.00 \\
ForecastBench       & -0.0448 & -0.0280 & -0.0628 & 1.00  & 1.00 \\
\bottomrule
\end{tabular}
\end{table}

\newpage

\begin{figure}[!htbp]
    \centering
    \includegraphics[width = \textwidth]{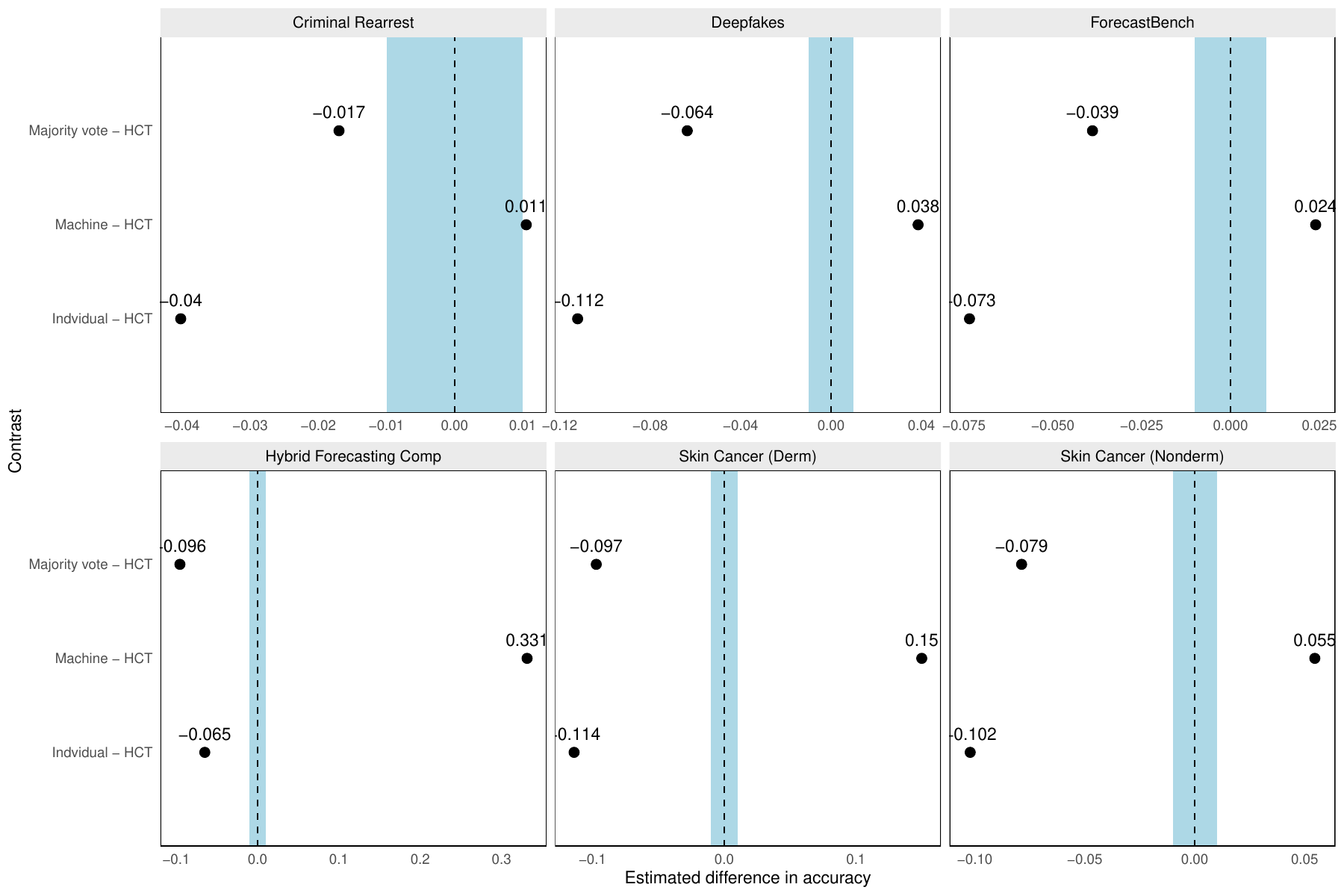}
    \caption{Cross-validation results per dataset, showing the difference in accuracy between the hybrid confirmation tree (HCT) and the majority vote, the machine, and one human decision maker. Smaller x-axis values indicate a greater performance benefit of the hybrid confirmation tree in comparison to the other strategy. Blue areas show the region of practical equivalence of $\pm$ 1 percent accuracy. The 95$\%$ HDI around the estimates are too small to be visible. The estimated difference between the hybrid confirmation tree and the strategies closely matches the results if selecting the threshold that maximizes the hybrid confirmation tree accuracy in-sample (Figure \ref{fig3}). See Supplementary Table \ref{tab:posterior} for an in-sample comparison between the hybrid confirmation tree and the majority vote, again, preserving the direction and magnitude with which the hybrid confirmation tree is more accurate.}
    \label{si_cv_marginals}
\end{figure}

\begin{figure}[!htbp]
    \centering
    \includegraphics[width = \textwidth]{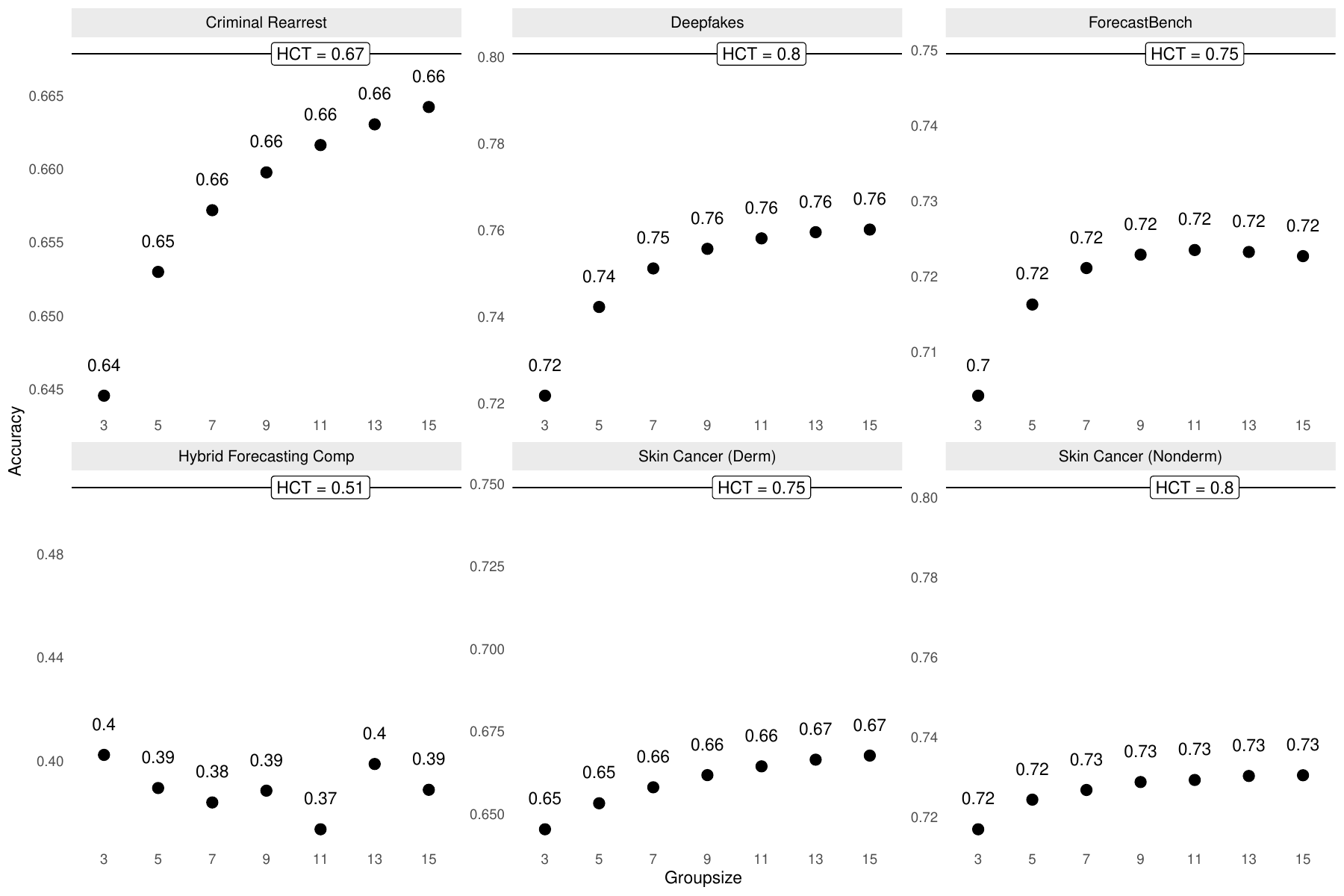}
    \caption{Accuracy of the majority vote (y-axis) for increasing group sizes (x-axis) as compared to the accuracy of the hybrid confirmation tree (solid line).}
    \label{si_larger_crowds}
\end{figure}

\newpage

\begin{figure}[!htbp]
    \centering
    \includegraphics[width = \textwidth]{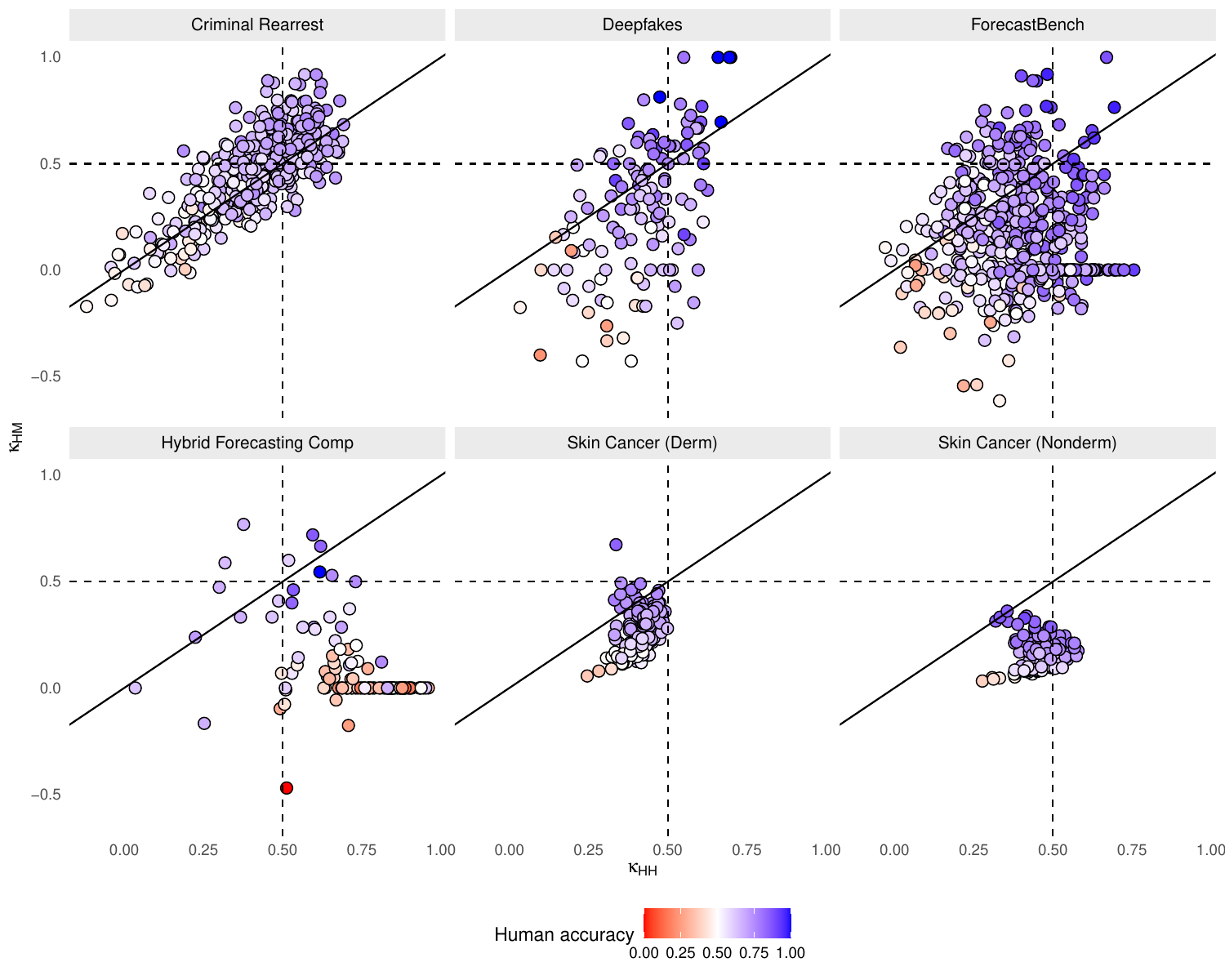}
    \caption{Human--human and human--machine correlation. Each dot corresponds to one human decision maker. The x-axis presents the correlation among humans ($\kappa_{HH}$); the y-axis presents the correlation between individual humans and machine choices ($\kappa_{HM}$). Color coding indicates the individual human's accuracy over all cases.}
    \label{si_kappa_empirical}
\end{figure}

\newpage

\begin{figure}[!htbp]
    \centering
    \includegraphics[width = \textwidth]{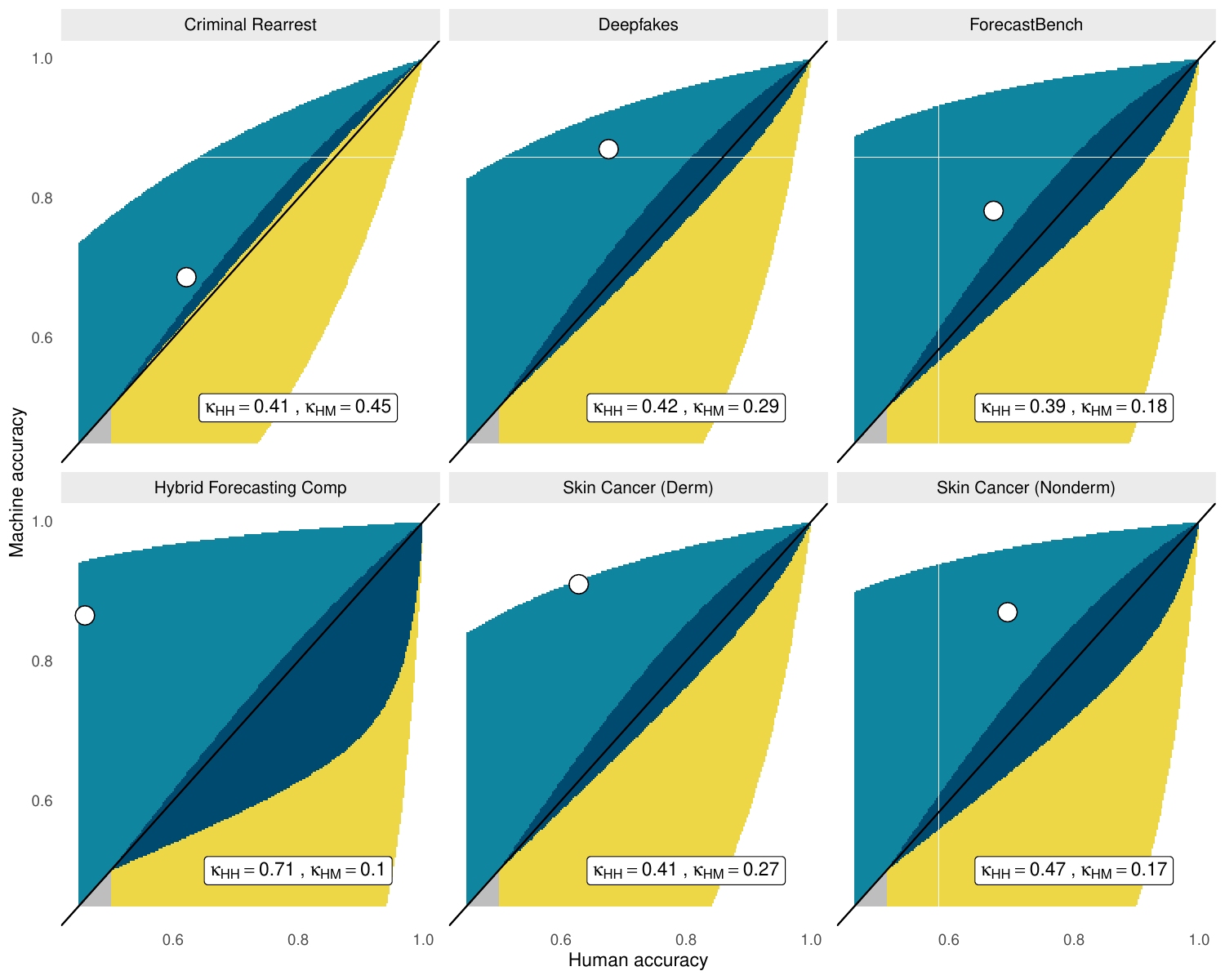}
    \caption{Empirical and analytical comparison of the hybrid confirmation tree, single humans, the machine, and a three-person majority vote as a function of human accuracy and machine accuracy and their respective correlations. Colors indicate regions where the hybrid confirmation tree outperforms the human majority vote (blue), the machine (yellow), and both the human majority vote and the machine (dark blue), and where it performs worse than the human majority and a single human vote (grey). These regions are based on the data set average estimated human--human $\kappa_{HH}$ and human--machine $\kappa_{HM}$ (bottom-right in facet). White dots indicate the human and machine accuracy averages per dataset for the threshold values that maximize hybrid confirmation tree accuracy.}
    \label{si_theory_data}
\end{figure}

\newpage

\begin{figure}[!h]
    \centering
    \includegraphics[width = \textwidth]{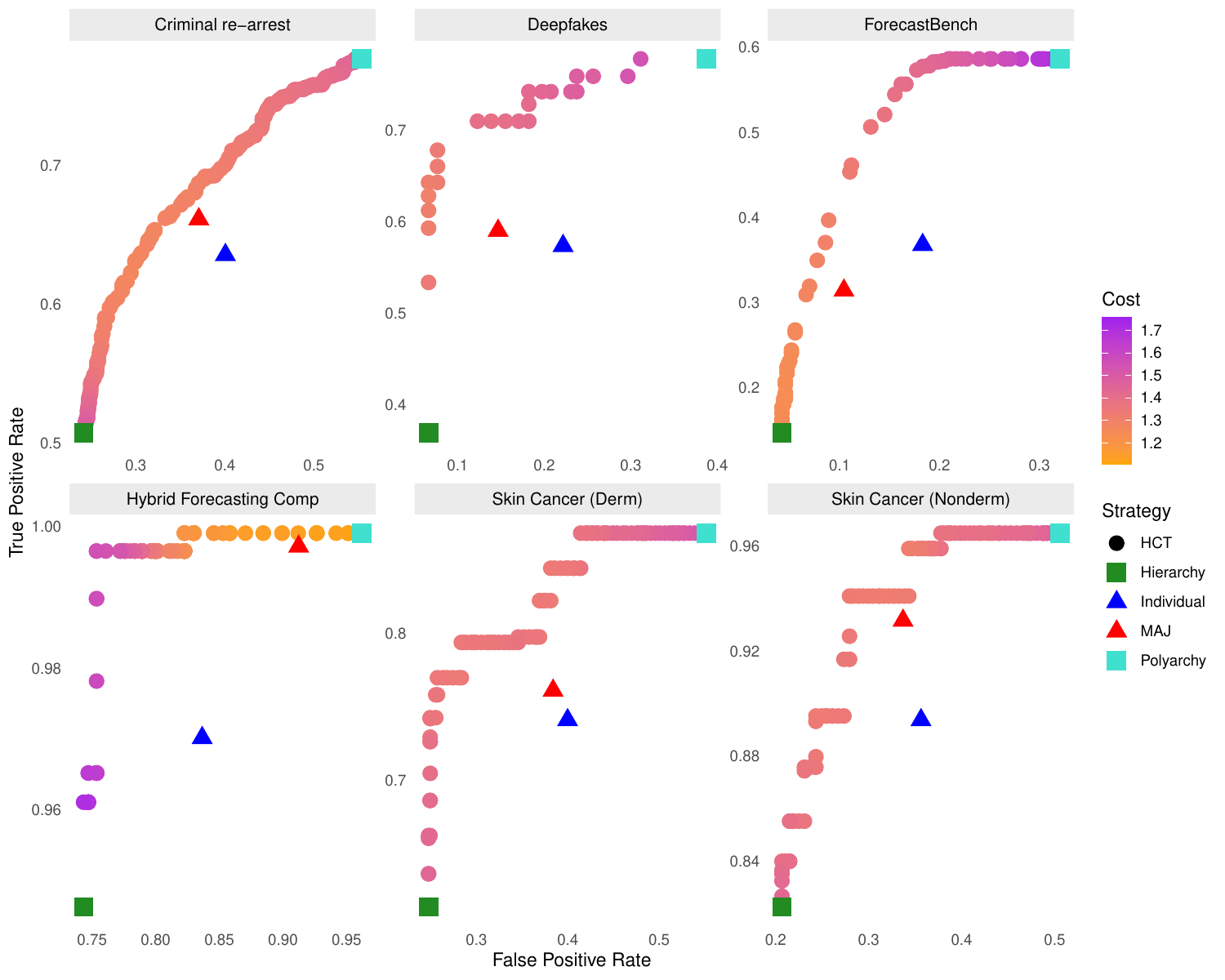}
    \caption{Empirical results: Flexible error trade-offs across domains. Results show the hybrid confirmation tree (HCT) true positive rate (y-axis) and false positive rate (x-axis) for every possible threshold between 0 and 1 that varies how easily a probabilistic prediction of the machine learning model in the HCT is counted as a positive or negative class (colored dots). The color coding corresponds to the cost of the hybrid confirmation tree measured in how many humans needed to be consulted for a final decision on average. The performance of the average human decision maker (blue triangle), the human majority vote (red triangle), the hierarchy (green square) and the polyarchy (turquoise square) are also presented.}
    \label{si_zoom}
\end{figure}

\newpage

\newpage

\begin{figure}[!t]
    \centering
    \includegraphics[width = \textwidth]{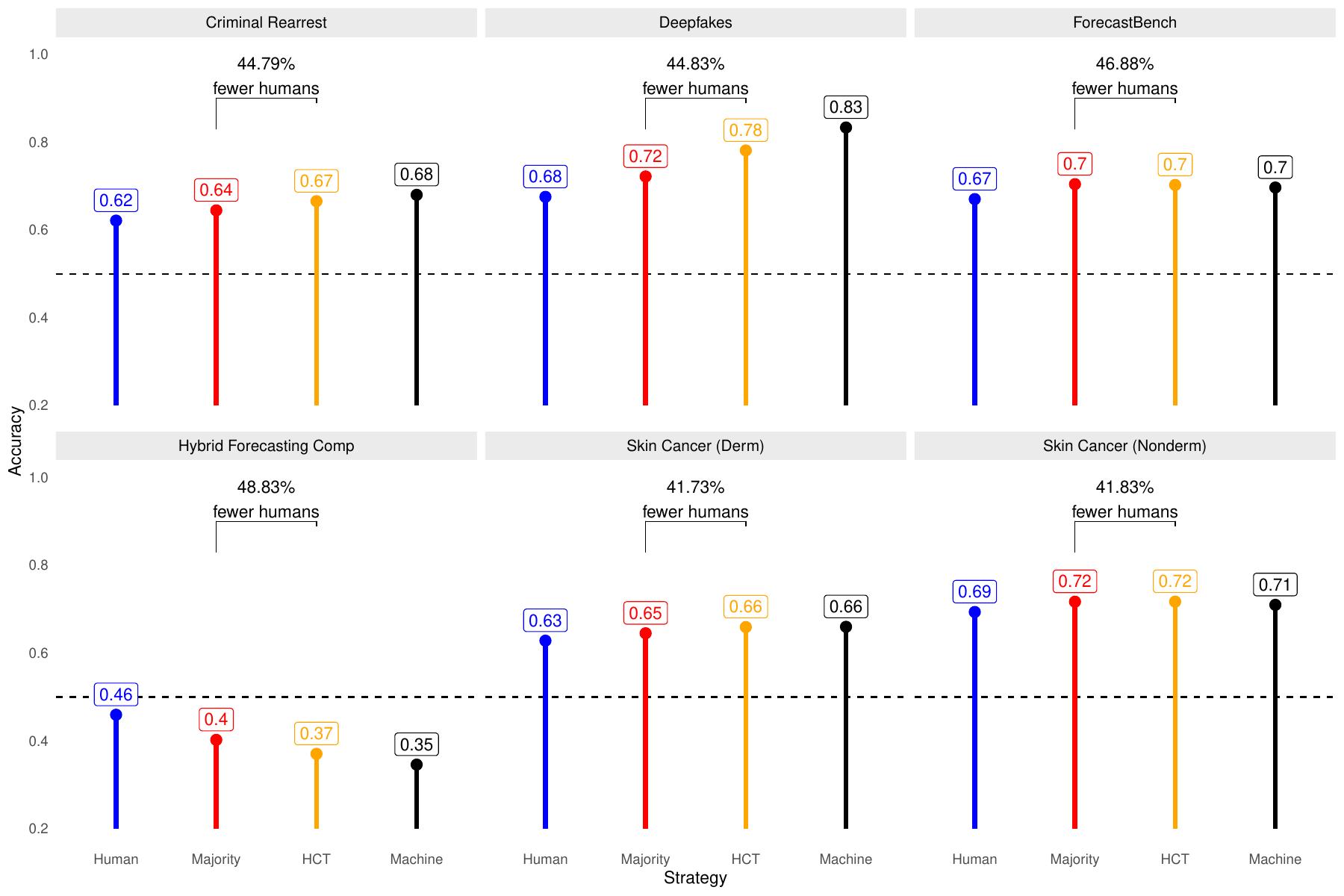}
    \caption{Comparing the accuracy of single humans, three-person majority vote, the hybrid confirmation tree (HCT) and the machine alone per domain. Results are generated by selecting the threshold setting that minimizes hybrid confirmation tree cost. Brackets indicate the relative reduction in human decision makers of the hybrid confirmation tree compared to the human majority vote.}
    \label{si_cost}
\end{figure}

\newpage

\begin{figure}[!t]
    \centering
    \includegraphics[width = \textwidth]{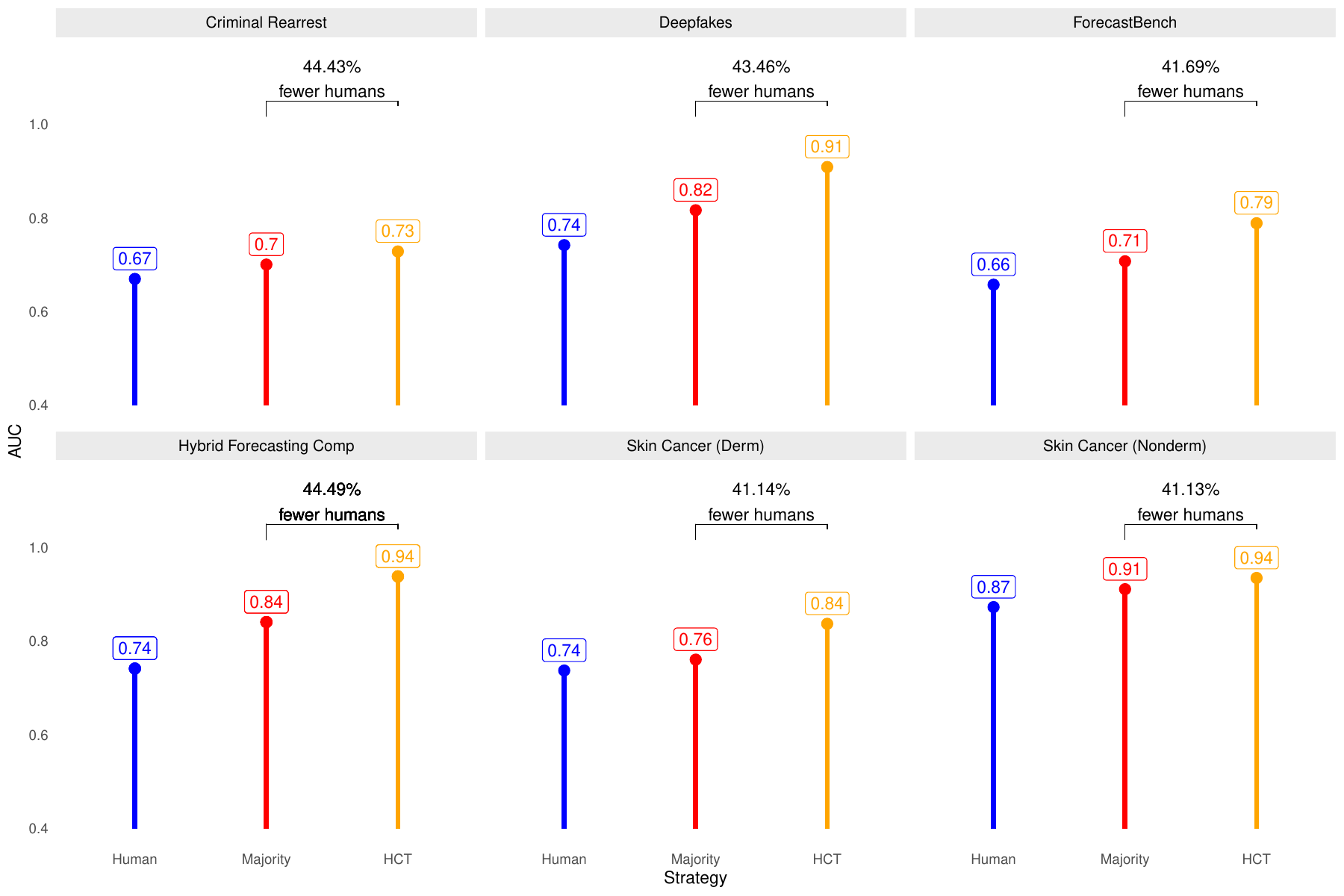}
    \caption{Comparison of single-point AUC estimates for the hybrid confirmation tree (HCT), single humans, and three-person majority vote across domains. The y-axis displays an AUC estimate derived from the specific True Positive Rate (TPR) and False Positive Rate (FPR) at a single operating point, calculated using the binormal formula $
\mathrm{AUC} = \Phi\bigl(\bigl[\Phi^{-1}(\mathrm{TPR}) - \Phi^{-1}(\mathrm{FPR})\bigr] / \sqrt{2}\bigr)
$. For the HCT, results correspond to the threshold setting that maximizes this single-point estimate. We omit the machine AUC in this figure because it is calculated conventionally (as the area integrated under the full ROC curve) and is therefore not directly comparable to these single-point estimates. Brackets indicate the relative reduction in human decision makers of the hybrid confirmation tree compared to the human majority vote.}
    \label{si_auc}
\end{figure}

\end{document}